\documentclass[preprint]{elsarticle}
 \pdfoutput=1
\usepackage{graphicx}
\usepackage{amssymb}    
\usepackage{epstopdf}

\usepackage{lscape}
\usepackage{epsfig}
\usepackage{fullpage}
\usepackage{graphicx}
\usepackage{fancyhdr}
\usepackage{amsmath,amssymb,xspace}
\usepackage{tabularx}

\newfont{\bbb}{msbm10 scaled\magstep1}
\newtheorem{theo}{Theorem}[section]

\newtheorem{prop}{Proposition}[section]

\newtheorem{rem}{Remark}[section]

\let \leq \leqslant
\let \geq \geqslant

\let \epsilon \varepsilon

{ \par \medskip \par
  \noindent \textit{\textbf{Demonstration\/}} : }{\null \hfill $\Box$ \par }
 
\newcommand{\R} {\ensuremath{\mathbb{R}}}

\newcommand{\N} {\ensuremath{\mathbb{N}}}
\newcommand{\C} {\ensuremath{\mathbb{C}}}

\journal{Journal of Computational Physics}

\begin{document}

\begin{frontmatter}



\title{%
Galerkin method for unsplit 3-D Dirac equation using atomically$/$kinetically balanced B-spline basis}

\author[inrs,crm]{F. Fillion-Gourdeau}
\ead{filliong@emt.inrs.ca}

\author[carl,crm]{E. Lorin}
\ead{elorin@math.carleton.ca}

\author[sher,crm]{A. D. Bandrauk}
\ead{andre.bandrauk@usherbrooke.ca}


\address[inrs]{Universit\'{e} du Qu\'{e}bec, INRS-\'{E}nergie, Mat\'{e}riaux et T\'{e}l\'{e}communications, Varennes, Canada, J3X~1S2}
\address[crm]{Centre de Recherches Math\'{e}matiques, Universit\'{e} de Montr\'{e}al, Montr\'{e}al, Canada, H3T~1J4}
\address[carl]{School of Mathematics and Statistics, Carleton University, Ottawa, Canada, K1S 5B6}
\address[sher]{Laboratoire de chimie th\'{e}orique, Facult\'{e} des Sciences, Universit\'{e} de Sherbrooke, Sherbrooke, Canada, J1K 2R1}

\date{\today}

\begin{abstract}
A Galerkin method is developed to solve the time-dependent Dirac equation in prolate spheroidal coordinates for an electron-molecular two-center system. The initial state is evaluated from a variational principle using a kinetic/atomic balanced basis, which allows for an efficient and accurate determination of the Dirac spectrum and eigenfunctions. B-spline basis functions are used to obtain high accuracy. This numerical method is used to compute the energy spectrum of the two-center problem and then the evolution of eigenstate wavefunctions in an external electromagnetic field.
\end{abstract}


\end{frontmatter}

\section{Introduction}

In the last few decades, there has been a surge of interest for the numerical solution of the Dirac equation in many areas of physics and chemistry, motivated mostly by new advances in computational architectures and numerical methods, which allow to tackle complex physical problems. One specific field that has benefited from these advances is laser-matter interaction where it is now possible to reach laser intensities of $10^{20}$ W/cm$^2$ \cite{RevModPhys.78.309} and higher in laboratories. The theoretical description of matter subject to such intense radiation can only be described by relativistic quantum mechanics which requires solutions of the Dirac equation \cite{RevModPhys.84.1177}. Traditionally however, the Dirac equation has been studied mostly in the context of relativistic heavy ion collisions, where the search for positron production from Uranium nuclear collisions is one of the main impetus \cite{PhysRevA.24.103}.  

Various numerical methods have been developed to solve the relativistic equation as analytical approaches are often challenging and only perturbative. Among the most popular approach is the operator splitting method, where the Dirac operator is separated into a set of simpler equations. Each of these resulting equations can then be solved by resorting to well-known and accurate numerical schemes. As there exists many possible decompositions of the Dirac operator, there also exists many variations of the operator splitting method. It is often combined with spectral methods whereby the kinetic operator is solved by the Fourier Transform methods while the mass and potential terms, being local operators in ``real space'', can be dealt with by accurate approximations of time-ordered exponentials. This technique has been used in \cite{15,p4,p10,p9} for the Dirac equation and in \cite{16,17} for the coupled Maxwell-Dirac equation that includes the interaction and the backreaction on the electromagnetic field. Another possible decomposition of the Dirac Hamiltonian was given in \cite{Succi1993327,Lorin_Bandrauk,FillionGourdeau20121403} using Alternate Direction Iteration. In this case, the spin is kept aligned with the direction of propagation at each step (using a specific rotation in spinor space) such that simple analytical solutions can be found using the method of characteristics. The resulting scheme, sometimes called ``Quantum Lattice Boltzmann'', can be parallelized very efficiently. It can also be adapted to treat the cylindrical coordinate case \cite{2013arXiv1303.3781F} and nonlinear Dirac equations \cite{PhysRevLett.111.160602}.

Although these approaches are very powerful and have very interesting properties, they are inefficient for finding the initial state of the system in a confining potential with bound and continuum states. Within the operator splitting method, these states are usually determined from a relativistic variant of the Feit-Fleck method \cite{15,p4,p10,2013arXiv1303.3781F}. The latter allows for the computation of the spectrum and the determination of bound states from a filtering technique on the time evolution of the wavefunction, thus having a very slow convergence. Therefore, these methods are impractical for problems in Quantum Electrodynamics requiring sums over all states of the spectrum.  For these reasons, other approaches have been considered. One possibility is the use of the mapped Fourier grid, which allows to evaluate both the spectrum and the time evolution of the wavefunction \cite{0305-4470-38-14-007,PhysRevA.78.062711}. One problem however with this scheme is the appearance of spurious states, which are unphysical states created during the discrete evolution process. Direct approaches, where the Dirac operator is discretized without splitting, have also been attempted. For instance, implicit finite difference schemes can be found in  \cite{PhysRevA.40.5548,PhysRevA.40.5559,p3,PhysRevC.71.024904} while an explicit scheme is in \cite{0022-3700-16-11-017}.  

Conversely, there exists very powerful schemes to solve the time-independent Dirac equation, based on variational methods and basis set expansion. The most important issue in this case is the variational collapse \cite{QUA:QUA560250112}, which is related to the fact that the spectrum of the Dirac equation is not bounded from below (or above). This induces spurious states in the spectrum obtained from the usual Rayleigh-Ritz variational method. This phenomenon is also called spectral pollution \cite{lewin}. There has been several (successful) attempts to solve this problem and there now exists two main lines of development:
\begin{enumerate}
 \item New variational principles
 \item Balance principles 
\end{enumerate} 
The first case corresponds to a modification of the usual Rayleigh-Ritz minmax principle. This was first investigated by Talman \cite{PhysRevLett.57.1091} and was generalized in \cite{dolbeault_2000,Dolbeault2000208}. This has led to numerical methods free of spurious states, but which requires the solution of a nonlinear eigenvalue problem (see \cite{Dolbeault2003} for instance). The latter usually requires an iteration method and thus, necessitates a lot of computation time. The second case corresponds to a modification of the basis function expansion such that spinor components are related in some ways. This was first introduced as an empirical rule to get rid of spurious states \cite{QUA:QUA560250112} and was then analyzed by comparing with the non-relativistic results \cite{1402-4896-36-3-013,PhysRevA.62.022508}. However, the rigorous analysis of these methods is fairly recent \cite{lewin}. There exists three well-known variations of the balance principle:
\begin{enumerate}
\item Kinetically balanced basis function \cite{10.1063/1.447865}
\item Atomic balanced basis function \cite{QUA:QUA560400816}
\item Dual kinetic balanced basis function \cite{PhysRevLett.93.130405}
\end{enumerate}
In each of them, a different relation is imposed between basis functions of the large and small spinor components. In this work, the atomic as well as kinetic balance will be used to compute the initial state (Cauchy data) for the time-dependent Galerkin method. 

The Galerkin method has been applied to the Dirac equation using different coordinate systems and basis sets \cite{PhysRevA.85.033411,PhysRevA.86.052705,FroeseFischer2009879}, for both time-dependent and time-independent cases. In this article, we develop numerical schemes to study the two-center problem in an external electromagnetic field. This system has also been investigated extensively, mostly in connection with heavy ion collisions and heavy ion spectroscopy. The static case can be found in \cite{PhysRevA.48.2700,Laaksonen1984485,0305-4470-38-14-007,FFG,Kullie2004215,Kullie1999307,
0953-4075-43-23-235207,Dage1994469,1402-4896-36-3-004,fillion2011aa} (an analytical approximation can be found in \cite{Muller19735}) but less is known for the dynamic case \cite{PhysRevA.78.062711,PhysRevA.86.052705,0022-3700-16-11-017,PhysRevA.89.012514}. The main goal of this article is to give a variant of these methods, based on atomic balance and B-spline basis sets.

This article is separated as follows. In Section \ref{DE}, we describe the Dirac equation studied in this article. Section \ref{TIDE} is devoted to the derivation and analysis of the Galerkin solver for the time-independent Dirac Hamiltonian. In Section \ref{TDDE}, we derive from the Time Independent Dirac Equation (TIDE) solver, a Time Dependent Dirac Equation (TDDE) version. Some mathematical properties of the derived schemes are also proposed in this section. Some important details of the numerical implementation are given in Section \ref{sec:impl}, along with some performance benchmarks. The numerical results are presented for TIDE and TDDE in Section \ref{NUM}.  We finally conclude in Section \ref{conc}.

\section{Dirac Equation}\label{DE}

The Dirac equation is a quantum wave equation that describes the relativistic dynamics of spin-$\frac{1}{2}$ particles (fermions) such as the electron. In this setting, the particle under consideration is characterized by a four-component spinor  
\begin{eqnarray*}
\Psi = [\phi,\chi]^T \in C^1\big(0,T;L^{2}(\R^3,\C^4)\big),
\end{eqnarray*}
for some positive time $T$. The bispinors $\phi,\chi \in C^1\big(0,T;L^{2}(\R^3,\C^2)\big)$ are usually, respectively called the large and small components. In the time-independent case, where we consider an interaction with a nucleus defined by a static external Coulomb potential $V_{c}$, the wavefunction obeys the following Dirac equation\footnote{All the calculations will be performed in atomic units (a.u.) where $m=1$, $\hbar =1$ and $c=1/\alpha $ where we take $\alpha \approx 1/137.035999679$ as the fine structure constant. In all the equations however, we are keeping the mass explicitly, allowing to switch easily from atomic to natural units.}:
\begin{eqnarray*}
i\partial_t \Psi = H_0\Psi, \;\;\mbox{with} \;\; H_0 \equiv  c \boldsymbol{\alpha} \cdot \mathbf{p} +  mc^{2} \beta + V_c(x) \mathbb{I}_{4} ,
\end{eqnarray*}
where $\boldsymbol{\alpha} = (\alpha_x,\alpha_y,\alpha_z)$ are the Dirac matrices, $H_0$ is the Hamiltonian operator, $\mathbf{p} = -i \nabla$ is the momentum operator, $c$ is the light velocity, $m$ is the electron mass, and $\Psi$ is the four component spinor. The matrix structure is given by $\boldsymbol{\alpha}$ and $\beta$ in $M_{4}(\C)$:
\begin{eqnarray}
 \alpha_{i} = \left[
\begin{array}{cc}
 0 & \sigma_{i} \\
\sigma_{i} & 0
\end{array} \right]
\; \mbox{and} \;
 \beta = \left[
\begin{array}{cc}
 \mathbb{I}_{2} & 0 \\
0 & -\mathbb{I}_{2}
\end{array} \right].
\end{eqnarray}
where $\sigma_{i}$ are the usual Pauli matrices. The latter are
\begin{eqnarray}
 \sigma_{x} = \left[
\begin{array}{cc}
 0 & 1 \\
1 & 0
\end{array} \right]
\; \mbox{,} \;
\sigma_{y} = \left[
\begin{array}{cc}
 0 & -i \\
i & 0
\end{array} \right]
\; \mbox{and} \;
 \sigma_{z} = \left[
\begin{array}{cc}
 1 & 0 \\
0 & -1
\end{array} \right].
\end{eqnarray}

We consider now the relativistic spin-$\frac{1}{2}$ quantum particle subject to a classical electromagnetic field $({\bf A},V) \in C^2(\R^3\times \R_+,\R^4)$. We will assume that the electromagnetic field is given at any time and that the back-reaction of the particle on the electromagnetic field is neglected. Therefore, Maxwell's equations are not solved numerically and we parametrize the electromagnetic field by an analytical form, given below (we refer to \cite{Lorin_Bandrauk,17,16} for a full Maxwell-Dirac equation solver based on another approach). The equation we consider is then:
\begin{eqnarray*}
i\partial_t \Psi = H\Psi, \, H = {\boldsymbol{\alpha}}\cdot\big(-ic{\nabla}-e{\bf A}\big) + mc^2\beta + \big(V_c(x)+V(t,x)\big)\mathbb{I}_{4},
\end{eqnarray*}
where the electromagnetic field was added by the minimal coupling prescription, which guarantees a gauge invariant formulation. In explicit calculations however, a specific gauge is chosen: we choose the Coulomb gauge $\nabla \cdot \mathbf{A} = 0$ such that the Coulomb law can be used to describe the static charged nuclei. We also set $V=0$ such that the laser field is characterized by the vectorial potential. 

This last equation gives a consistent description of bound electrons in molecules in the Born-Oppenheimer approximation, i.e. when the nuclei are fixed in space and included in the potential term $V_c$. This is a valid approximation when the mass of the nucleus is much larger than the mass of the electron thus neglecting momentum exchange between photons, electrons and nuclei, which will always be the case for the systems considered in this study.

\subsection{Dirac equation for the two-center system, prolate spheroidal coordinates and boundary conditions}

We focus on the simple electron molecular two-center system where we consider two nuclei described by the Coulomb potential, as 
\begin{eqnarray}
V_{c} = -\frac{Z_{1}e}{\sqrt{x^{2}+y^{2}+(z-R)^{2}}} -\frac{Z_{2}e}{\sqrt{x^{2}+y^{2}+(z+R)^{2}}},
\end{eqnarray} 
where $Z_{1,2}$ are the nuclear charges, $R$ is the internuclear distance and $x,y,z$ are Cartesian coordinates.

To treat this system, it is convenient first to consider cylindrical coordinates where
\begin{eqnarray}
x = r \cos(\theta), \, y = r \sin(\theta),
\end{eqnarray}
where $r = \sqrt{x^{2}+y^{2}}$ is the radial distance and $\theta = \tan^{-1}(y/x)$ is the azimuthal angle. Assuming that the Dirac equation has an azimuthal symmetry, which occurs when the electrodynamic potential does not depend on $\theta$, it is then possible to reduce the number of dimensions from 3 to 2 by separation of variables. The $\theta$-dependence can be factorized by using the following ansatz for the four-spinor with cylindrical symmetry \cite{0305-4470-16-9-024,Kullie2004215}:
\begin{eqnarray}
\label{eq:ansatz}
 \Psi (\mathbf{x},t)  = \left[
\begin{array}{c}
 \psi_{1}(t,r,z) e^{i\mu_{1}\theta} \\
\psi_{2}(t,r,z) e^{i\mu_{2}\theta} \\
\psi_{3}(t,r,z) e^{i\mu_{1}\theta} \\
\psi_{4}(t,r,z) e^{i\mu_{2}\theta}
\end{array} 
\right] ,
\end{eqnarray}
where $\mu_{1,2} := j_{z} \mp 1/2$ and where $j_{z}$ is the angular momentum projection on the $z$-axis (it can take one of the values $j_{z} = \cdots,-\frac{5}{2},-\frac{3}{2},-\frac{1}{2},\frac{1}{2},\frac{3}{2},\frac{5}{2},\cdots$). Substituting in the Dirac equation leads to
\begin{eqnarray}
i\partial_{t}\psi(t,r,z)  &=& \biggl\{ \alpha_{x} \biggl[ -ic \partial_{r} -ic\frac{1}{2r} - eA_{r}(t,r,z) \biggr] + \alpha_{y}  \biggl[ c\frac{j_{z}}{r} - eA_{\theta}(t,r,z) \biggr] \nonumber \\
&&  + \alpha_{z} \biggl[ -ic \partial_{z} - eA_{z}(t,r,z) \biggr]  + \beta  m c^{2} + eV_{c}(r,z) \biggr\}  \psi(t,r,z).
\label{eq:dirac_cyl3}
\end{eqnarray}
Then, by using the symmetry of the coordinate transformation and by assuming that the wave function is regular enough, it is demonstrated in \cite{2013arXiv1303.3781F} that the wave function can be written as 
\begin{eqnarray}
\label{eq:fac1}
\psi_{1}(t,r,z) &=& r^{|\mu_{1}|} \varphi_{1}(t,r^{2},z) ,\\
\label{eq:fac2}
\psi_{2}(t,r,z) &=& r^{|\mu_{2}|} \varphi_{2}(t,r^{2},z) ,\\
\label{eq:fac3}
\psi_{3}(t,r,z) &=& r^{|\mu_{1}|} \varphi_{3}(t,r^{2},z) ,\\
\label{eq:fac4}
\psi_{4}(t,r,z) &=& r^{|\mu_{2}|} \varphi_{4}(t,r^{2},z) ,
\end{eqnarray}
where $\varphi$ admits a Taylor expansion in $r^{2}$ around $r=0$.
Therefore, the boundary conditions at $r=0$ on $\psi$ is a Robin condition which depends on the value of $\mu_{1}$ and $\mu_{2}$ \cite{2013arXiv1303.3781F}. These boundary conditions will be included in the numerical scheme with the addition of a prefactor in basis functions \cite{Kullie2004215,FFG}.

For the two-center problem, it is known that prolate spheroidal coordinates yield more accurate results in both the relativistic and non-relativistic cases. Moreover, in these coordinates, the nuclei are positioned at the corners of the domain, facilitating the numerical implementation. For these reasons, we now turn to these coordinates. The prolate spheroidal coordinates which are related to cylindrical coordinates as follows 
\begin{eqnarray}
\label{eq:r_vs_pro}
 r &=& R \left[ (\xi^{2} - 1)(1-\eta^{2}) \right]^{\frac{1}{2}}  , \\ 
 \label{eq:z_vs_pro}
 z &=& R \xi \eta ,
\end{eqnarray}
where $\xi \in [1,\infty)$, $\eta \in [-1,1]$ and $\theta = [0,2\pi]$ (azimuthal angle). This choice is particularly attractive when dealing with a two center potential. To obtain the Dirac equation in these coordinates, one simply uses the mapping in Eqs. \eqref{eq:r_vs_pro} and \eqref{eq:z_vs_pro} along with the derivatives

\begin{eqnarray}
\label{eq:deri_r}
 \partial_{r} &=& \frac{\sqrt{(\xi^{2}-1)(1-\eta^{2})}}{R(\xi^{2}-\eta^{2})} \left[ \xi \partial_{\xi} - \eta \partial_{\eta} \right] ,\\
\label{eq:deri_z}
\partial_{z} &=& \frac{(\xi^{2}-1)}{R(\xi^{2}-\eta^{2})} \eta \partial_{\xi} + \frac{(1-\eta^{2})}{R(\xi^{2}-\eta^{2})} \xi \partial_{\eta}.
\end{eqnarray}

\subsection{Dirac equation in the time-independent case: Cauchy data}

The goal of this paper is to accurately solve the TDDE for particles subject to a classical electromagnetic field. Prior to this, we first determine the initial data of the Cauchy problem which is naturally chosen as the ground (or any bound) state of the Dirac Hamiltonian. We are then require to solve the TIDE:
\begin{eqnarray}
\label{eq:dirac}
H_0 \psi (x) = E \psi(x) .
\end{eqnarray}
It is convenient to write the four-spinor as $\psi (x) \equiv \left[ \phi(x) , \chi(x) \right]^{\rm T} \in L^{2}(\R^3,\C^4)$ where $\phi(x)$ and $\chi(x)$ are the large and small components, respectively. The eigenvalue problem \eqref{eq:dirac} reduces explicitly to
\begin{eqnarray}
\label{eq:dir_mat}
 \left[
\begin{array}{cc}
 V_c(x) + mc^{2} & R_0 \\
R_0 & V_c(x) - mc^{2}
\end{array} \right]
 \left[
\begin{array}{c}
 \phi(x) \\
 \chi(x)
\end{array} \right]
=
E
\left[
\begin{array}{c}
 \phi(x) \\
 \chi(x)
\end{array} \right]
\end{eqnarray}
where 
\begin{eqnarray}
R_{0}:=\sigma_{x} \biggl[ -ic \partial_{r} -ic\frac{1}{2r}\biggr] + \sigma_{y}  c\frac{j_{z}}{r} -ic \sigma_{z} \partial_{z} .
\end{eqnarray}
Equation \eqref{eq:dir_mat} is equivalent to
\begin{eqnarray}
 R_0 \chi(x) &=& [E-mc^{2} - V_c(x)] \phi(x) \\
R_0 \phi(x) &=& [E+mc^{2} - V_c(x)] \chi(x)
\end{eqnarray}
which is the common starting point for the numerical method that follows. The small component can then be written in terms of the large component yielding
\begin{eqnarray}
\label{eq:sm_comp}
\chi(x) = \frac{R_0}{E+mc^{2} - V_c(x)}  \phi(x) 
\end{eqnarray}
%
This relation will be important for the analysis that follows concerning balance principles.

\section{Time Independent Dirac Equation Solver}\label{TIDE}

In this section, the numerical method used to compute the TIDE is described. As stated above, this is required to obtain the initial state of the time evolution of the wavefunction. The latter will be given in the next section.

\subsection{Rayleigh-Ritz method}

The Rayleigh-Ritz method is based on a variational principle which allows to estimate the eigenvalues of a given operator. These eigenvalues can be characterized by the following variational principle:
\begin{eqnarray}
\label{eq:ray_ritz_func}
\bar{H}_0= \cfrac{\langle \psi | H_0 |\psi \rangle_{L^2(\R^3,\C^4)}}{ \langle \psi | \psi \rangle_{L^2(\R^3,\C^4)}}, 
\end{eqnarray}
which is nothing but the usual Rayleigh-Ritz coefficient. Finding the eigenvalue by this minimization procedure is equivalent to finding the stationary point of the functional
\begin{eqnarray}\label{MINE}
\mathcal{E}[\psi] = \langle \psi | H_0 |\psi \rangle_{L^2(\R^3,\C^4)} - E \langle \psi | \psi \rangle_{L^2(\R^3,\C^4)}
\end{eqnarray}
where the energy becomes a Lagrangian multiplier. This form will be used in the following to convert the basis set expansion into a generalized eigenvalue problem. It is well-known that the convergence of this method depends on the fact that the spectrum is bounded from below. This is not the case for the Dirac operator, owing to the presence of the negative energy states and this may induce spurious states in the spectrum. This is discussed in the next section.

\subsection{About spectral pollution}
This section is an non-exhaustive summary of some key results about spectral pollution for the approximate Dirac Hamiltonian, constructed using Galerkin's techniques and balance principles. In this approach, one still uses the Rayleigh-Ritz variational principle, but with a different set of basis functions which approximates the relation between small and large spinor components given in Eq. \eqref{eq:sm_comp}.

A spurious state can be defined rigorously as follows. Notations, proofs and additional results can be found in \cite{lewin}. We just summarize some key ideas of this very strong and quite technical work. We consider an operator $A$ of domain $D(A) \subseteq H$, where $H$ is a Hilbert space. An eigenvalue $\lambda \in \R$ is said spurious for Operator $A$ if there exists a sequence of finite dimensional vector spaces $(V_n)_n \subseteq D(A)$ and  $V_{n} \subseteq V_{n+1}$ such that 
\begin{itemize}
\item  $\overline{\cup_{n \geq 1}V_n}^{D(A)} = D(A)$
\item $\lim_n\hbox{dist}\big(\lambda,\sigma(A_{|V_n})\big)=0$
\item $\lambda \notin \sigma(A)$
\end{itemize}
The last item emphasizes that the eigenvalue is not in the spectrum of the operator and thus, non-physical. Therefore, a strategy has to be developed to eliminate these states. One possibility is the use of balanced basis functions which are defined in the following way and for which we summarize some properties from \cite{lewin}.

The $4$-component spinor is split via a projector operator $P : H\rightarrow H$, defined by $P[\phi,\chi]^T=[\phi,0]^T$ for $\phi$ and $\chi$ in $L^2(\R^3,\C^2)$. A balanced operator $L: D(L) \subseteq PH \rightarrow (1-P)H$ is defined from $P$ as follows
\begin{itemize}
\item $L$ is 1-1
\item $D(L)\oplus LD(L)$ is a core of $A$
\end{itemize}
The notion of a spurious eigenvalue of Operator $A$ associated to a projector $P$ and balanced operator $L$ can finally be defined. Assuming that there exists a sequence of finite dimensional vectors spaces $(V_n^+)_n$ such that $V_{n}^+ \subseteq D(L)$ and $V_n^+ \subseteq V_{n+1}^+$ and
\begin{itemize}
\item $\overline{\cup_n\big(V_n^+ \oplus LV_n^+\big)}^{D(A)}=D(A)$
\item $\lim_n\hbox{dist}\big(\lambda,\sigma(A_{|V^+_n\oplus LV_n^+})\big)=0$
\item $\lambda \notin \sigma(A)$
\end{itemize}
The corresponding spurious spectrum is denoted Spu$\big(A,P,L\big)$. In this framework, the kinetically balanced operator is defined by 
\begin{eqnarray*}
L_{KB} = \cfrac{1}{2mc^2}\boldsymbol{\alpha}\cdot{\bf p}
\end{eqnarray*}
and the atomic balanced operator is defined
\begin{eqnarray*}
L_{AB} = \cfrac{1}{2mc^2-V_c}\boldsymbol{\alpha}\cdot{\bf p}
\end{eqnarray*}
Two of the main theorems of \cite{lewin} state
\begin{theo}\label{theoKB}
Assuming that $V_c$ is of the form $V_c(x)=-\kappa |x|^{-1}$ for $\kappa \in (0,3/2)$ (which includes Coulomb potentials) then
\begin{eqnarray*}
\overline{\hbox{Spu}\big(H_0+V_c,P,L_{KB}\Big)} = [-1,1]
\end{eqnarray*}
\end{theo}
\begin{theo}
Assuming that $V_c$ is such that $V_c(x) \geq -\kappa|x|^{-1}$ for $\kappa \in (0,3/2)$ with $\sup(V_c)<2$, $(2-V_c)^{-2}\nabla V_c \in L^{\infty}(\R^3)$ and $\max(V_c,0) \in L^p(\R^3)$ with $p>3$ and $V_c(x)\rightarrow_{\infty} 0$, then
\begin{eqnarray*}
\overline{\hbox{Spu}\big(H_0+V_c,P,L_{AB}\Big)} = [-1,-1+\sup(V_c)]
\end{eqnarray*}
In particular for Coulomb potentials, the spurious spectrum is always empty.
\end{theo}
According to these results for Coulomb potentials, spectral pollution can be generated with kinetically balanced bases, but not with atomic balanced bases. However, it should be noted that for a given basis set, spurious may as well not appear. As we are interested in the two-center system with Coulomb potential, from the spuriousity perspective it is preferably to use the atomic balance basis set. Notice that the numerical tests performed below have not exhibited any spurious state with the kinetically balanced operators (which is however not in contradiction with Theorem \ref{theoKB}).

\subsection{Variational method and balanced basis set}

The basis which is chosen to expand $\psi$ is a B-spline basis constructed as follows. First, following \cite{FFG}, we expand the small component $\phi$ as:

\begin{eqnarray}
\label{eq:basis_1}
 \phi_{1,2}(\xi,\eta) &=& \sum_{n=1}^{N} a_{n}^{(1,2)}B^{(1,2)}_{n}(\xi,\eta) 
\end{eqnarray}
where $a_{n}^{(1,2)}$ are the coefficients of the basis expansion and $B^{(1,2)}_{n}(\xi,\eta)$ are the basis functions, for components 1 and 2 respectively, expressed in the prolate spheroidal coordinate system $\xi,\eta$ described in Eqs. \eqref{eq:r_vs_pro} and \eqref{eq:z_vs_pro}. The basis function can then be written as the tensor product of B-spline functions $b_{i}^{k}(x)$ of order $k$ as
\begin{eqnarray}
\label{eq:basis_func_def}
B^{(1,2)}_{n}(\xi,\eta) =  G^{(1,2)}(\xi,\eta)b_{i}^{k_{\xi}}(\xi)b_{j}^{k_{\eta}}(\eta)
\end{eqnarray}
where $n = [i,j] \in \mathbb{Z}^{2}$, $i \in [1,n_{\xi}]$ and $j \in [1,n_{\eta}]$. Some properties of B-splines are recalled in the next section.  
An overall factor is used to account for angular momentum dependence \cite{Kullie2004215,Kullie1999307,FFG}. It is defined by

%
%
%
\begin{eqnarray}
G^{(1,2)}(\xi,\eta) = r^{|\mu_{1,2}|},
\end{eqnarray}
consistent with the boundary conditions in Eqs. \eqref{eq:fac1} to \eqref{eq:fac4}.
%
%

Using the atomic balance approach, the lower spinor components are then expanded as follows, in the atomic balance case:
\begin{eqnarray}\label{ABE}
 \chi = \frac{R_0}{2mc^2-V_c} \left(
\begin{array}{c}
 \sum_{n=1}^{N} c_{n}^{(1)}B^{(1)}_{n} \\
\sum_{n=1}^{N} c_{n}^{(2)}B^{(2)}_{n}
\end{array}
\right)
\end{eqnarray}
and as follows in the kinetic balance case
\begin{eqnarray}
 \chi = \frac{R_0}{2mc^2} \left(
\begin{array}{c}
 \sum_{n=1}^{N} c_{n}^{(1)}B^{(1)}_{n} \\
\sum_{n=1}^{N} c_{n}^{(2)}B^{(2)}_{n}
\end{array}
\right)
\end{eqnarray}
{\it In the following, the presentation is done in the atomic balance framework. We refer to \cite{FFG} or Remark \ref{rem1}, for the kinetic balance framework.}  In prolate spheroidal coordinates, \eqref{ABE} becomes
\begin{eqnarray}
\label{chi}
\chi_{1}(\xi,\eta) &=& \cfrac{ic}{2mc^{2}-V_c}\sum_{n=1}^{N} 
\biggl\{ c_{n}^{(2)}\left[ - \partial_{r} - \frac{\mu_{2}}{r}  \right]  B^{(2)}_{n} -   c_{n}^{(1)}\partial_{z} B^{(1)}_{n} \biggr\}, \\
\chi_{2}(\xi,\eta) &=& \cfrac{ic}{2mc^{2}-V_c} \sum_{n=1}^{N} \biggl\{ c_{n}^{(1)} \left[ - \partial_{r} + \frac{\mu_{1}}{r}  \right]  B^{(1)}_{n} +  c_{n}^{(2)} \partial_{z}B^{(2)}_{n} \biggr\}.
\end{eqnarray}
These formulae should be understood as $r := r(\xi,\eta)$ and $z:=z(\xi,\eta)$ where the relations are given in Eqs. \eqref{eq:r_vs_pro} and \eqref{eq:z_vs_pro} for coordinates and in Eqs. \eqref{eq:deri_r} and \eqref{eq:deri_z} for derivatives. 

%
%
The very first step to solve our Cauchy problem is to determine the initial condition. In physical situations, it is often chosen as the ground state for the considered system of particles. We then have to solve an eigenvalue problem: $H_0 \psi_0 = E_0\psi_0$, where $H_0$ is the field-free Dirac Hamiltonian. The variational formulation corresponds to finding stationary points of the functional
\begin{eqnarray}
\label{eq:ray_ritz}
\mathcal{E}[\psi] &=& \langle \phi | ( V_{c} + mc^{2} ) \phi \rangle_{L^2(\R^3,\C^2)} + \langle R_0 \phi |  \chi \rangle_{L^2(\R^3,\C^2)} \nonumber \\
& &  + \langle \chi | R_0 \phi \rangle_{L^2(\R^3,\C^2)} + \langle \chi |(V_{c} - mc^{2}) \chi \rangle_{L^2(\R^3,\C^2)}  \nonumber \\
& &- E \left [\langle \phi | \phi \rangle_{L^2(\R^3,\C^2)} - \langle \chi |\chi \rangle_{L^2(\R^3,\C^2)} \right],
\end{eqnarray}
which is just an explicit way of writing the well-known Rayleigh-Ritz functional equation in Eq. \eqref{eq:ray_ritz_func}. Integration by part was used to write the second term in a convenient form. The notation $\langle \cdot|\cdot \rangle_{L^2(\R^3,\C^2)}$ stands for the Hermitian inner product on $L^2(\R^3,\C^2)$. In the following, we define 2 operators $\mathbf{C}$ and $\mathbf{S}$ by
\begin{eqnarray}
 C[\psi] &=& \int_{\mathbb{R}^{3}} \left[ mc^{2} + V_{c} \right] |\phi|^{2} + (R_0 \phi|\chi) + (\chi|R_0 \phi) + [V_{c} - mc^{2}] |\chi|^{2} \\
S[\psi] &=& \int_{\mathbb{R}^{3}} |\phi|^{2} + |\chi|^{2}.
\end{eqnarray}
Here, the product $( \cdot | \cdot )$ is just the spinor product\footnote{For $\Xi$ a two-component spinor, it is defined as $(\Xi | \Xi) = \Xi^{*}_{1}\Xi_{1} + \Xi^{*}_{2}\Xi_{2}$.}.
Using the atomically balanced bases, as described above, and finding the stationary points of $\mathcal{E}$ by setting
\begin{eqnarray}
\cfrac{\partial \mathcal{E}}{\partial a_{i}^{(1)*}} = 0 , \cfrac{\partial \mathcal{E}}{\partial a_{i}^{(2)*}} = 0 , \cfrac{\partial \mathcal{E}}{\partial c_{i}^{(1)*}} = 0, \cfrac{\partial \mathcal{E}}{\partial c_{i}^{(2)*}} = 0
\end{eqnarray}
for $i \in \{1,\cdots,N\}$, we obtain the following discrete generalized eigenvalue problem:
\begin{eqnarray}
\label{eq:gen_eig}
 \mathbf{C} \mathbf{a} = E \mathbf{S} \mathbf{a}
\end{eqnarray}
where $\mathbf{a}= [ a^{(1)}_{1},\cdots,a^{(1)}_{N},a^{(2)}_{1},\cdots,a^{(2)}_{N},c^{(1)}_{1},\cdots,c^{(1)}_{N},c^{(2)}_{1},\cdots,c^{(2)}_{N}]$ and
\begin{eqnarray}
 \mathbf{C}  = \left[
\begin{array}{cccc}
 \mathbf{C}^{(1)}_{11} &0&\mathbf{C}^{(3)}_{11}&\mathbf{C}^{(3)}_{12} \\
0 & \mathbf{C}^{(1)}_{22} &\mathbf{C}^{(3)}_{21}& \mathbf{C}^{(3)}_{22} \\
\mathbf{C}^{(3) \rm T}_{11} & \mathbf{C}^{(3) \rm T}_{21} & \mathbf{C}^{(2)}_{11} & \mathbf{C}^{(2)}_{12} \\
\mathbf{C}^{(3) \rm T}_{12} & \mathbf{C}^{(3) \rm T}_{22} & \mathbf{C}^{(2) \rm T}_{11} & \mathbf{C}^{(2)}_{22}
\end{array}
\right], \mathbf{S} =&\left[
\begin{array}{cccc}
 \mathbf{S}^{(1)}_{11} &0& 0 & 0 \\
0 & \mathbf{S}^{(1)}_{22} & 0 & 0 \\
0 & 0 & \mathbf{S}^{(2)}_{11} & \mathbf{S}^{(2)}_{12} \\
0 & 0 & \mathbf{S}^{(2) \rm T}_{12} & \mathbf{S}^{(2)}_{22}
\end{array}
\right] 
\end{eqnarray}
The elements of these matrices are defined by:
\begin{eqnarray}
\label{eq:C_mat_1}
\bigl[ \mathbf{C}^{(1)}_{11} \bigr]_{ij} & = &  \int d^{3}x \biggl\{ (V_c+mc^{2})B^{(1)}_{i}B^{(1)}_{j} \biggr\} \\
\bigl[ \mathbf{C}^{(1)}_{22} \bigr]_{ij} & = &  \int d^{3}x \biggl\{ (V_c+mc^{2})B^{(2)}_{i}B^{(2)}_{j} \biggr\} 
\end{eqnarray}
\begin{eqnarray}
\bigl[ \mathbf{C}^{(2)}_{11} \bigr]_{ij} & = &  \int d^{3}x \biggl\{  (\partial_{z}B^{(1)}_{i})(\partial_{z}B^{(1)}_{j}) 
+(\partial_{r}B^{(1)}_{i})(\partial_{r}B^{(1)}_{j}) \nonumber \\
&&\quad \quad \quad + \frac{\mu_{1}^{2}}{r^{2}} B^{(1)}_{i}B^{(1)}_{j}  - \frac{\mu_{1}}{r} B^{(1)}_{i}(\partial_{r}B^{(1)}_{j}) \nonumber \\
&&\quad \quad \quad - \frac{\mu_{1}}{r} (\partial_{r}B^{(1)}_{i})B^{(1)}_{j}  \biggr\} \cfrac{(V_c-mc^{2})c^2}{(2mc^2-V_c)^2} \\
\bigl[ \mathbf{C}^{(2)}_{22} \bigr]_{ij} & = &   \int d^{3}x \biggl\{  (\partial_{z}B^{(2)}_{i})(\partial_{z}B^{(2)}_{j}) 
+(\partial_{r}B^{(2)}_{i})(\partial_{r}B^{(2)}_{j}) \nonumber \\
&&\quad \quad \quad + \frac{\mu_{2}^{2}}{r^{2}} B^{(2)}_{i}B^{(2)}_{j} + \frac{\mu_{2}}{r} B^{(2)}_{i}(\partial_{r}B^{(2)}_{j}) \nonumber \\
&&\quad \quad \quad + \frac{\mu_{2}}{r} (\partial_{r}B^{(2)}_{i})B^{(2)}_{j}  \biggr\}\cfrac{(V_c-mc^{2})c^2}{(2mc^2-V_c)^2}\\
\bigl[\mathbf{C}^{(2)}_{12}\bigr]_{ij} & = & \int d^{3}x \biggl\{  (\partial_{z}B^{(1)}_{i})(\partial_{r}B^{(2)}_{j}) 
+ \frac{\mu_{1}}{r} B^{(1)}_{i}(\partial_{z}B^{(2)}_{j})  \nonumber \\ 
&&\quad \quad \quad  -(\partial_{r}B^{(1)}_{i})(\partial_{z}B^{(2)}_{j})
+ \frac{\mu_{2}}{r} (\partial_{z}B^{(1)}_{i})B^{(2)}_{j} \biggr\} \nonumber \\
&&\quad \quad \quad  \times \cfrac{(V_c-mc^{2})c^2}{(2mc^2-V_c)^2}
\end{eqnarray}
\begin{eqnarray}
\bigl[\mathbf{C}^{(3)}_{11}\bigr]_{ij} & = &  \int d^{3}x \biggl\{  (\partial_{z}B^{(1)}_{i})(\partial_{z}B^{(1)}_{j}) 
+(\partial_{r}B^{(1)}_{i})(\partial_{r}B^{(1)}_{j}) \nonumber \\
&&\quad \quad \quad + \frac{\mu_{1}^{2}}{r^{2}} B^{(1)}_{i}B^{(1)}_{j}  - \frac{\mu_{1}}{r} B^{(1)}_{i}(\partial_{r}B^{(1)}_{j}) \nonumber \\
&&\quad \quad \quad - \frac{\mu_{1}}{r} (\partial_{r}B^{(1)}_{i})B^{(1)}_{j}  \biggr\}\frac{c^2}{2mc^2-V_c}  \\
\bigl[\mathbf{C}^{(3)}_{22}\bigr]_{ij} & = &   \int d^{3}x \biggl\{  (\partial_{z}B^{(2)}_{i})(\partial_{z}B^{(2)}_{j}) 
+(\partial_{r}B^{(2)}_{i})(\partial_{r}B^{(2)}_{j}) \nonumber \\
&&\quad \quad \quad + \frac{\mu_{2}^{2}}{r^{2}} B^{(2)}_{i}B^{(2)}_{j} + \frac{\mu_{2}}{r} B^{(2)}_{i}(\partial_{r}B^{(2)}_{j}) \nonumber \\
&&\quad \quad \quad + \frac{\mu_{2}}{r} (\partial_{r}B^{(2)}_{i})B^{(2)}_{j}  \biggr\} \cfrac{c^2}{2mc^2-V_c}  \\
\bigl[\mathbf{C}^{(3)}_{12}\bigr]_{ij} & = & \int d^{3}x \biggl\{  (\partial_{z}B^{(1)}_{i})(\partial_{r}B^{(2)}_{j}) 
+ \frac{\mu_{1}}{r} B^{(1)}_{i}(\partial_{z}B^{(2)}_{j})  \nonumber \\ 
&&\quad \quad \quad  -(\partial_{r}B^{(1)}_{i})(\partial_{z}B^{(2)}_{j})
+ \frac{\mu_{2}}{r} (\partial_{z}B^{(1)}_{i})B^{(2)}_{j} \biggr\}  \cfrac{c^2}{2mc^2-V_c}
\end{eqnarray}
and
 \begin{eqnarray}
 \label{eq:explicit_nRR2}
 \bigl[\mathbf{S}^{(1)}_{11}\bigr]_{ij} & = &  \int d^{3}x \biggl\{ B^{(1)}_{i}B^{(1)}_{j} \biggr\}  = \bigl[\mathbf{S}^{(2)}_{11}\bigr]_{ij}\\
 \bigl[\mathbf{S}^{(1)}_{22}\bigr]_{ij} & = &  \int d^{3}x \biggl\{ B^{(2)}_{i}B^{(2)}_{j} \biggr\} = \bigl[\mathbf{S}^{(2)}_{22}\bigr]_{ij} 
 \end{eqnarray}
\begin{eqnarray} \label{eq:explicit_nRR2b}
\bigl[ \mathbf{S}^{(2)}_{11} \bigr]_{ij} & = &  \int d^{3}x \biggl\{  (\partial_{z}B^{(1)}_{i})(\partial_{z}B^{(1)}_{j}) 
+(\partial_{r}B^{(1)}_{i})(\partial_{r}B^{(1)}_{j}) \nonumber \\
&&\quad \quad \quad + \frac{\mu_{1}^{2}}{r^{2}} B^{(1)}_{i}B^{(1)}_{j}  - \frac{\mu_{1}}{r} B^{(1)}_{i}(\partial_{r}B^{(1)}_{j}) \nonumber \\
&&\quad \quad \quad - \frac{\mu_{1}}{r} (\partial_{r}B^{(1)}_{i})B^{(1)}_{j}  \biggr\} \cfrac{c^2}{(2mc^2-V_c)^2} \\
\bigl[ \mathbf{S}^{(2)}_{22} \bigr]_{ij} & = &   \int d^{3}x \biggl\{  (\partial_{z}B^{(2)}_{i})(\partial_{z}B^{(2)}_{j}) 
+(\partial_{r}B^{(2)}_{i})(\partial_{r}B^{(2)}_{j}) \nonumber \\
&&\quad \quad \quad + \frac{\mu_{2}^{2}}{r^{2}} B^{(2)}_{i}B^{(2)}_{j} + \frac{\mu_{2}}{r} B^{(2)}_{i}(\partial_{r}B^{(2)}_{j}) \nonumber \\
&&\quad \quad \quad + \frac{\mu_{2}}{r} (\partial_{r}B^{(2)}_{i})B^{(2)}_{j}  \biggr\}\cfrac{c^2}{(2mc^2-V_c)^2}\\
\label{eq:S_mat_end}
\bigl[\mathbf{S}^{(2)}_{12}\bigr]_{ij} & = & \int d^{3}x \biggl\{  (\partial_{z}B^{(1)}_{i})(\partial_{r}B^{(2)}_{j}) 
+ \frac{\mu_{1}}{r} B^{(1)}_{i}(\partial_{z}B^{(2)}_{j})  \nonumber \\ 
&&\quad \quad \quad  -(\partial_{r}B^{(1)}_{i})(\partial_{z}B^{(2)}_{j})
+ \frac{\mu_{2}}{r} (\partial_{z}B^{(1)}_{i})B^{(2)}_{j} \biggr\} \nonumber \\
&&\quad \quad \quad  \times \cfrac{c^2}{(2mc^2-V_c)^2}
\end{eqnarray}
 These last expressions can then be rewritten in prolate spheroidal coordinates. In practice, the eigenvalue problem in Eq. \eqref{eq:gen_eig} is solved by a standard eigensolver for sparse matrices.
The integration measure is given in prolate spheroidal coordinates by
\begin{eqnarray}
\label{eq:meas_pro_coord}
 d^{3}x = R^{3}(\xi^{2}-\eta^{2})d\xi d\eta d \theta .
\end{eqnarray}
\begin{rem}\label{rem1}
 Note that Matrices $\mathbf{C}$, $\mathbf{D}$ and $\mathbf{S}$ are very similar to those obtained with kinetically balanced bases \cite{FFG}. The difference comes from the presence in atomic balance of the $V_c$-term in the denominators $\cfrac{c^2}{(2mc^2-V_c)^2}$, the latter is absent when using a kinetically balanced basis. 
\end{rem}
Although atomic balance is, from the variational collapse viewpoint, more attractive than kinetic balance, the presence of $V_c$ (atomic balance) is seen to be source of numerical discrepancy of the overall convergence rate, and special treatment is then necessary to tackle this additional difficulty. 
\subsection{Some basic facts about B-splines}
We recall some basic facts about B-splines. We refer for instance to \cite{KC} for details. First B-splines are fully determined by their order $k_{\xi,\eta}$ and knot vector using the following iterative formula
\begin{eqnarray}
b_{i}^{k}(x) = \frac{x-t_{i}}{t_{i+k-1} - t_{i}}b^{k-1}_{i}(x) + \frac{t_{i+k} -x}{t_{i+k}-t_{i+1}} b^{k-1}_{i+1}(x) 
\end{eqnarray}
with initial conditions
\begin{eqnarray}
 b_{i}^{1}(x) = 1 \;\mbox{for} \; t_{i} \leq x < t_{i+1} \; \mbox{and} \; b_{i}^{1} = 0 \; \mbox{otherwise}
\end{eqnarray}
where $t_{i}$'s are knots coordinates. The number of knots, also referred as breaking points, at a given coordinates essentially determines the regularity conditions at that point: the number of knots points should be maximal at singular points (at the Coulomb singularity position for instance) to allow for a discontinuous-like behavior. As in \cite{FFG}, throughout this work, the knot vectors are given by the sequences 
\begin{eqnarray}
1 = \xi_{1} = \cdots = \xi_{k_{\xi}} < \xi_{k_{\xi}+1} < \cdots< \xi_{n_{\xi}+1} = \cdots = \xi_{n_{\xi}+k_{\xi}} = \xi_{\rm max} \\
-1 = \eta_{1} = \cdots = \eta_{k_{\eta}} < \eta_{k_{\eta}+1} < \cdots< \eta_{n_{\eta}+1} = \cdots = \eta_{n_{\eta}+k_{\eta}} = 1 
\end{eqnarray}
Here, $n_{\xi,\eta} $ are the number of spline functions in $\xi$ and $\eta$ coordinates respectively.
\\
\\
Considering $f \in C^r([a,b])$ ($r \in \N$) the distance between $f$ and the space of B-splines $S_n^k$ of degree $k$, with $r<k<n$, is given by 
\begin{eqnarray}
\label{err1}
\hbox{dist}(f,S_n^k) = \inf_{g \in S_n^k}\|f-g\| \leq k^r\max_{-k \leq i \leq n+1}|x_i-x_{i+1}|^r \|f^{(r)}\|_{\infty}
\end{eqnarray} 
As a consequence, assuming that the solution to the Dirac equation is regular enough, we can expect a very good accuracy with high order B-splines. The methods developed here are Galerkin's methods and therefore, require the numerical evaluation of several integrals. 
In this work, Gauss' quadrature methods will be used to approximate the integrals constituting the stiffness, mass matrices. We recall that 
\begin{eqnarray*}
\Big|\int_{-1}^1f(x)dx  -  \sum_{i=1}^n \omega_i f(x_i) \Big| \leq  \cfrac{2^{2n}(n!)^4}{2n!(2n+1)!R^{2n}}\max_{|z|=R}|f(z)|
\end{eqnarray*}
with $(x_i)_i$ roots of Legendre's polynomials and $(\omega_i)_i$ its weights..

\section{Time Dependent Dirac Equation Solver}\label{TDDE}
\subsection{Finite element method for TDDE}\label{Galer}
The Cauchy problem we now consider is:
\begin{eqnarray}
\label{eq1}
i\partial_t \psi = H \psi, \, (t,x)\in \R_+\times \R^3, \qquad \psi(0,x) = \psi_0(x), x\in \R^3
\end{eqnarray}
The initial data $\psi_0$ is taken as a state for the time-independent Dirac operator, that is an eigenfunction associated with one of the eigenvalues of $H_0$. Now in the field dependent case the TDDE can be rewritten:
\begin{eqnarray*}
i\partial_t\left[
\begin{array}{c}
 \phi(t,x) \\
 \chi(t,x)
\end{array} \right]
= \left[
\begin{array}{cc}
 V_c(x) + mc^{2} & R \\
R & V_c(x) - mc^{2}
\end{array} \right]
 \left[
\begin{array}{c}
 \phi(t,x) \\
 \chi(t,x)
\end{array} \right]
\end{eqnarray*}
with $R := R_{0} - e \boldsymbol{\sigma} \cdot \mathbf{A} $, that is
\begin{eqnarray*}
i\partial_t\left[
\begin{array}{c}
 \phi(t,x) \\
 \chi(t,x)
\end{array} \right] = 
\left[
\begin{array}{cc}
 V_c(x) + mc^{2} & R_0 \\
R_0 & V_c(x) - mc^{2}
\end{array} \right]
 \left[
\begin{array}{c}
 \phi(t,x) \\
 \chi(t,x)
\end{array} \right]
+ \left[
\begin{array}{cc}
0 & -e\boldsymbol{\sigma}\cdot{\bf A} \\
-e\boldsymbol{\sigma}\cdot{\bf A}& 0
\end{array}\right]
 \left[
\begin{array}{c}
 \phi(t,x) \\
 \chi(t,x)
\end{array} \right]
\end{eqnarray*}
Where ${\bf A}$ is the vector potential corresponding to to some field ${\bf E}$, ${\bf E} = -\partial_t{\bf A}$ and ${\bf B} = \nabla \times {\bf A}$, in the Coulomb gauge, where $V=0$.

To get a Galerkin method from the preceding equation, we have to project on basis functions. To perform this procedure, we introduce a basis spline $\mathcal{B}$ defined by the atomically balanced procedure described above, which give the $j$'th basis function spinor as
\begin{eqnarray}
\mathcal{B}_{j} := 
\begin{bmatrix}
\phi_{1,j} \\
\phi_{2,j} \\
\chi_{1,j} \\
\chi_{2,j}
\end{bmatrix}
=
\begin{bmatrix}
B^{(1)}_{j} \\
B^{(2)}_{j} \\
\cfrac{ic}{2mc^{2}-V_c} \biggl\{ \left[ - \partial_{r} - \frac{\mu_{2}}{r}  \right]  B^{(2)}_{j} -   \partial_{z} B^{(1)}_{j} \biggr\} \\
 \cfrac{ic}{2mc^{2}-V_c}  \biggl\{ \left[ - \partial_{r} + \frac{\mu_{1}}{r}  \right]  B^{(1)}_{j} +   \partial_{z}B^{(2)}_{j} \biggr\}
\end{bmatrix}
\end{eqnarray}
Then, the weak form of the Dirac equation is obtained as
\begin{eqnarray}
\langle \mathcal{B}_{j} |i \partial_{t} \psi \rangle_{L^2(\R^3,\C^4)} = \langle \mathcal{B}_{j} | H \psi \rangle_{L^2(\R^3,\C^4)}, \;\; \mbox{for} \;\; j \in \{1,\cdots,N\}
\end{eqnarray}
where as usual, the test functions were chosen as the basis function spinor $\mathcal{B}$. The last equation can be rewritten more explicitly as 
\begin{eqnarray}\label{FEM}
\langle \phi_{j} |i \partial_{t} \phi \rangle_{L^2(\R^3,\C^2)} + \langle \chi_{j} | i\partial_{t} \chi \rangle_{L^2(\R^3,\C^2)} &=&
\langle \phi_{j} |(V_{c} + mc^{2}) \phi \rangle_{L^2(\R^3,\C^2)} + \langle \chi_{j} |(V_{c}-mc^{2}) \chi \rangle_{L^2(\R^3,\C^2)} \nonumber \\
\nonumber \\
&&
+\langle \phi_{j} | R_{0} \chi \rangle_{L^2(\R^3,\C^2)} + \langle \chi_{j} |R_{0} \phi \rangle_{L^2(\R^3,\C^2)} \nonumber \\
\nonumber \\
&& - e\langle \phi_{j} |( \boldsymbol{\sigma}\cdot{\bf A} )\chi \rangle_{L^2(\R^3,\C^2)} -e \langle \chi_{j} | (\boldsymbol{\sigma}\cdot{\bf A}) \phi \rangle_{L^2(\R^3,\C^2)},
\end{eqnarray}
for $j \in \{1,\cdots,N\}$. These equations are then discretized by using a basis set expansion with time dependent coefficients as
\begin{eqnarray}
\phi_{1}(t,\xi,\eta) &=& \sum_{n=1}^{N} a^{(1)}_{n}(t) B^{(1)}_{n}(\xi,\eta) \\
\phi_{2}(t,\xi,\eta) &=& \sum_{n=1}^{N} a^{(2)}_{n}(t) B^{(2)}_{n}(\xi,\eta) \\
\chi_{1}(t,\xi,\eta) &=& \cfrac{ic}{2mc^{2}-V_c}\sum_{n=1}^{N} \biggl\{ c_{n}^{(2)}(t)\left[ - \partial_{r} - \frac{\mu_{2}}{r}  \right]  B^{(2)}_{n}(\xi,\eta) -   c_{n}^{(1)}(t)\partial_{z} B^{(1)}_{n}(\xi,\eta) \biggr\} \\
\chi_{2}(t,\xi,\eta) &=& \cfrac{ic}{2mc^{2}-V_c} \sum_{n=1}^{N} \biggl\{ c_{n}^{(1)}(t) \left[ - \partial_{r} + \frac{\mu_{1}}{r}  \right]  B^{(1)}_{n}(\xi,\eta) +  c_{n}^{(2)}(t) \partial_{z}B^{(2)}_{n}(\xi,\eta) \biggr\}
\end{eqnarray}
which are the atomically balanced basis functions described in the last section, but with time-dependent coefficients. We then arrive at the semi-discrete TDDE scheme, whereby the spatial discretisation is performed, which writes:
\begin{eqnarray*}
i{\bf S}\dot {\bf a}(t) = \big({\bf C} + {\bf D}(t)\big){\bf a}(t)
\end{eqnarray*}
with  $\mathbf{a}(t)= [ a^{(1)}_{1}(t),\cdots,a^{(1)}_{n}(t),a^{(2)}_{1}(t),\cdots,a^{(2)}_{n}(t),c^{(1)}_{1}(t),\cdots,c^{(1)}_{n}(t),c^{(2)}_{1}(t),\cdots,c^{(2)}_{n}(t)]$ is the time dependent unknown.

Possible time discretizations include:
\begin{itemize}
\item Explicit Euler scheme, which is nonunitary:
\begin{eqnarray*}
{\bf S}{\bf a}^{n+1} = {\bf S} {\bf a}^{n} -i \Delta t_n \big({\bf C} + {\bf D}^{n} \big){\bf a}^{n},
\end{eqnarray*}
where ${\bf a}^n = {\bf a}(t_n)$ for $n\in \N$.
\item Semi-implicit scheme (Crank-Nicolson scheme) which is unitary:
\begin{eqnarray}\label{EXP_SYS}
{\bf S}{\bf a}^{n+1} = {\bf S} {\bf a}^{n} -i \cfrac{\Delta t_n}{2} \big({\bf C} + {\bf D}^{n} \big){\bf a}^n - i\cfrac{\Delta t_n}{2} \big({\bf C} + {\bf D}^{n+1}  \big){\bf a}^{n+1}
\end{eqnarray}
or
\begin{eqnarray}\label{Heun}
{\bf S}{\bf a}^{n+1} = {\bf S} {\bf a}^{n} -i \cfrac{\Delta t_n}{2} \big({\bf C} + {\bf D}^{n} \big){\bf a}^n - i\cfrac{\Delta t_n}{2} \big({\bf C} + {\bf D}^{n}  \big){\bf a}^{n+1}
\end{eqnarray}
or more generally Runge-Kutta type schemes.
\item Simplectic integration schemes, such as:
\begin{eqnarray}
\mathbf{a}(t_{f}) & =& T\exp \left[ -i \int_{t_{i}}^{t_{f}} \mathbf{S}^{-1} \left(\mathbf{C} + \mathbf{D}(t) \right) \right] \mathbf{a}(t_{i}) \\
 & =& \exp \left[ -i  \mathbf{S}^{-1} \left(\mathbf{C} + \mathbf{D}(t_{i}+\delta t/2) \right) \right] \mathbf{a}(t_{i}) + O(\delta t^{3})
\end{eqnarray}
\end{itemize}

Matrices ${\bf S}$, ${\bf C}$, are identical to the ones defined in the time-independent case, in Eqs. \eqref{eq:C_mat_1} to \eqref{eq:S_mat_end}. The only time-dependent matrix is the one that includes the electromagnetic field $\mathbf{D}(t)$. It is obtained by discretizing in space the following terms of the weak functional:
\begin{eqnarray}
D(t) :=  - e\langle \phi_{j} |( \boldsymbol{\sigma}\cdot{\bf A} )\chi \rangle_{L^2(\R^3,\C^2)} -e \langle \chi_{j} | (\boldsymbol{\sigma}\cdot{\bf A}) \phi \rangle_{L^2(\R^3,\C^2)}
\end{eqnarray}
By using the basis expansion, it can be written as
\begin{eqnarray}
\mathbf{D} = 
\begin{bmatrix}
0 & 0 & \mathbf{D}^{(3)}_{11} & \mathbf{D}^{(3)}_{12} \\
0 & 0 & \mathbf{D}^{(3)}_{21} & \mathbf{D}^{(3)}_{22} \\
\mathbf{D}^{(4)}_{11} & \mathbf{D}^{(4)}_{12} &0&0 \\
\mathbf{D}^{(4)}_{21} & \mathbf{D}^{(4)}_{22} &0&0 
\end{bmatrix}.
\end{eqnarray}
The entries of this matrix are
\begin{eqnarray}
\bigl[\mathbf{D}^{(3)}_{11}\bigr]_{ij} &=& ie\int d^{3}x \biggl\{   A_{z} B_{i}^{(1)} (\partial_{z} B_{j}^{(1)})
- (A_{r}-iA_{\theta})  \left[ -B_{i}^{(1)}(\partial_{r}B_{j}^{(1)}) + \frac{\mu_{1}}{r}B_{i}^{(1)} B_{j}^{(1)}\right] \biggr\} \frac{c}{2mc^{2}-V_{c}} \\
\bigl[\mathbf{D}^{(3)}_{22}\bigr]_{ij} &=& ie\int d^{3}x \biggl\{   A_{z} B_{i}^{(2)} (\partial_{z} B_{j}^{(2)})
- (A_{r}+iA_{\theta})  \left[ -B_{i}^{(2)}(\partial_{r}B_{j}^{(2)}) - \frac{\mu_{2}}{r}B_{i}^{(2)} B_{j}^{(2)}\right] \biggr\} \frac{c}{2mc^{2}-V_{c}} \\
\bigl[\mathbf{D}^{(3)}_{12}\bigr]_{ij} &=& ie\int d^{3}x \biggl\{   -(A_{r}-iA_{\theta})B_{i}^{(1)} (\partial_{z} B_{j}^{(2)})
-  A_{z} \left[ -B_{i}^{(1)}(\partial_{r}B_{j}^{(2)}) - \frac{\mu_{2}}{r}B_{i}^{(1)} B_{j}^{(2)}\right] \biggr\} \frac{c}{2mc^{2}-V_{c}} \\
\bigl[\mathbf{D}^{(3)}_{21}\bigr]_{ij} &=& ie\int d^{3}x \biggl\{   (A_{r}+iA_{\theta})B_{i}^{(2)} (\partial_{z} B_{j}^{(1)})
+  A_{z} \left[ -B_{i}^{(2)}(\partial_{r}B_{j}^{(1)}) + \frac{\mu_{1}}{r}B_{i}^{(2)} B_{j}^{(1)}\right] \biggr\} \frac{c}{2mc^{2}-V_{c}}
\end{eqnarray}
and
\begin{eqnarray}
\bigl[\mathbf{D}^{(4)}_{11}\bigr]_{ij} &=& ie\int d^{3}x \biggl\{  - A_{z} (\partial_{z}B_{i}^{(1)})  B_{j}^{(1)}
+ (A_{r}+iA_{\theta})  \left[ -(\partial_{r}B_{i}^{(1)})B_{j}^{(1)} + \frac{\mu_{1}}{r}B_{i}^{(1)} B_{j}^{(1)}\right] \biggr\} \frac{c}{2mc^{2}-V_{c}} \\
\bigl[\mathbf{D}^{(4)}_{22}\bigr]_{ij} &=& ie\int d^{3}x \biggl\{  - A_{z} (\partial_{z}B_{i}^{(2)})  B_{j}^{(2)}
+ (A_{r}-iA_{\theta})  \left[ -(\partial_{r} B_{i}^{(2)}) B_{j}^{(2)} - \frac{\mu_{2}}{r}B_{i}^{(2)} B_{j}^{(2)}\right] \biggr\} \frac{c}{2mc^{2}-V_{c}} \\
\bigl[\mathbf{D}^{(4)}_{12}\bigr]_{ij} &=& ie\int d^{3}x \biggl\{  - (A_{r}-iA_{\theta})(\partial_{z}B_{i}^{(1)})  B_{j}^{(2)}
-  A_{z} \left[ -(\partial_{r}B_{i}^{(1)}) B_{j}^{(2)} + \frac{\mu_{1}}{r}B_{i}^{(1)} B_{j}^{(2)}\right] \biggr\} \frac{c}{2mc^{2}-V_{c}} \\
\bigl[\mathbf{D}^{(4)}_{21}\bigr]_{ij} &=& ie\int d^{3}x \biggl\{   (A_{r}+iA_{\theta}) (\partial_{z}B_{i}^{(2)}) B_{j}^{(1)}
+  A_{z} \left[ -(\partial_{r} B_{i}^{(2)}) B_{j}^{(1)} - \frac{\mu_{2}}{r}B_{i}^{(2)} B_{j}^{(1)}\right] \biggr\} \frac{c}{2mc^{2}-V_{c}}
\end{eqnarray}
Again prolate spheroidal coordinates are used to numerically evaluate these integrals.

\subsection{Mathematical properties}
The Galerkin method presented in Section \ref{Galer} has several nice and attractive mathematical features which are detailed in this section. These properties are valid, except when the opposite is specified, with kinetically and atomically balance bases. Recall first, that the use of prolate spheroidal coordinates leads to a very convenient position (for local mesh refinement) of the molecule nuclei at the corners of the domain.  \\
The first important result is related to the structure of Matrix $\mathbf{S}$ \eqref{eq:gen_eig} involved in the TDDE solver described in Section \eqref{Galer}.
\begin{prop}
Matrix $\mathbf{S}$ defined in \eqref{eq:gen_eig}, \eqref{eq:explicit_nRR2} and \eqref{eq:explicit_nRR2b} is a Hermitian matrix (real eigenvalues).
\end{prop} 
Although in principle,  $0$ can be an eigenvalue, it will be necessarily unique with atomically balanced basis (at least), as no spectral pollution is expected in that case, \cite{lewin}. 
\\
\\
In the sequel, we are interested in the consistency and stability of the time dependent solver. The field-free TDDE:
\begin{eqnarray*}
i\partial_t \psi = H_0 \psi, \qquad \psi(\cdot,0)= \phi_0(\cdot)
\end{eqnarray*}
where $H_0\phi_0 = E_0 \phi_0$, with $E_0$ the ground state energy, has the following exact solution $\phi_0(\cdot)\exp(-iE_0 t)$.  When this property is satisfied at the discrete level, up to the order of the time discretization, we will say that the TDDE solver is {\it consistent} with the eigenvalue solver. We have
\begin{prop}\label{const}
Assume that the time operator $\partial_t$ and variable, are discretized using i) an explicit Euler scheme or ii) a Crank-Nicolson scheme \eqref{Heun}, then the TDDE solver \eqref{Galer} is {\it consistent} with the eigenvalue solver \eqref{MINE}.  
\end{prop}
This simple property is very important from a practical point of view. In fact, this result can be extended to a large class of semi-discretization in time, but for the sake of simplicity we restrict the analysis to these two cases.\\
\noindent{\bf Proof.} The numerical ground state is constructed using the same atomically balanced basis and same mesh as the TDDE solver. Indeed in that case, ${\bf D}$ is identically zero, and the semi-discrete scheme becomes, for $t\geq 0$
\begin{eqnarray*}
i{\bf S}\dot {\bf a}(t) = {\bf C}{\bf a}(t)
\end{eqnarray*}
with, by assumption ${\bf a}(0)$, defined by ${\bf C}{\bf a}(0) = E_0 {\bf S}{\bf a}(0)$.\\
\\
Using the explicit Euler scheme, we get:
\begin{eqnarray*}
{\bf S}{\bf a}^1 = {\bf S} {\bf a}(0) -i \Delta t_0 {\bf C}{\bf a}(0) = {\bf S} {\bf a}(0) -i \Delta t_0 E_0{\bf S} {\bf a}(0), 
\end{eqnarray*}
which can be easily re-written as
\begin{eqnarray*}
\label{eq:one_timestep}
{\bf a}^1 = \big(1 -i \Delta t_0 E_0\big) {\bf a}(0).
\end{eqnarray*}
By induction and for time steps $\Delta t_l$, with $l\geq 0$, one obtains furthermore:
\begin{eqnarray*}
i{\bf S}\dot {\bf a}\Big(\sum_{l=0}^n\Delta t_l\Big) = {\bf C}{\bf a}\Big(\sum_{l=0}^n\Delta t_l\Big)
\end{eqnarray*}
Assuming that the solution at the $n$'th timestep is ${\bf a}^n = \Pi_{l=0}^{n-1}\big(1 -i \Delta t_l E_0\big) {\bf a}(0)$ and from
\begin{eqnarray*}
{\bf Sa}^{n+1}= \big({\bf S}-i\Delta t_n {\bf C}\big){\bf a}^n,
\end{eqnarray*}
we can obtain the solution at timestep $n+1$ by induction:
\begin{eqnarray*}
{\bf a}^{n+1} = \big(1 -i \Delta t_n E_0\big) {\bf a}^n = \Pi_{l=0}^{n}\big(1 -i \Delta t_l E_0\big) {\bf a}(0),
\end{eqnarray*}
where ${\bf a}^n = [ a^{(1),n}_{1},\cdots,a^{(1),n}_{N},a^{(2),n}_{1},\cdots,a^{(2),n}_{N},c^{(1),n}_{1},\cdots,c^{(1),n}_{N},c^{(2),n}_{1},\cdots,c^{(2),n}_{N}]$. This leads to the expected result that the discretized time evolution operator is
\begin{eqnarray*}
\Pi_{l=0}^{n}\big(1 -i \Delta t_l E_0\big) = 1-i E_0 \sum_{l=0}^n\Delta t_l + \mathcal{O}(n\Delta t_{\infty}^2) = \exp(-iE_0 \sum_{l=0}^n\Delta t_l) + \mathcal{O}(n\Delta t_{\infty}^2)
\end{eqnarray*}
where $\Delta t_{\infty} = \max_{0\leq j\leq n} \Delta t_j$. \\
\\
In the case of a (semi-implicit) Crank-Nicolson scheme, the same reasoning can be performed. For one time iteration, we get
\begin{eqnarray*}
{\bf S}{\bf a}^1 = {\bf S} {\bf a}(0) -i \cfrac{\Delta t_0}{2} {\bf C}{\bf a}(0) -i \cfrac{\Delta t_0}{2} {\bf C}{\bf a}^1.
\end{eqnarray*}
This is written as
\begin{eqnarray*}
{\bf S}{\bf a}^1 = {\bf S} {\bf a}(0) -i\cfrac{\Delta t_0}{2} E_0{\bf S} {\bf a}(0)-i \cfrac{\Delta t_0}{2} {\bf C}{\bf a}^1,
\end{eqnarray*}
from which we obtain
\begin{eqnarray*}
\big({\bf S} + i\cfrac{\Delta t_0}{2}{\bf C}\big) {\bf a}^1 =  {\bf S}\big(1 -i \cfrac{\Delta t_0}{2} E_0\big) {\bf a}(0).
\end{eqnarray*}
Then, we deduce an explicit form for the time iteration given by
\begin{eqnarray*}
 {\bf a}^1 =  \big(1 -i \cfrac{\Delta t_0}{2} E_0\big)\big({\bf S} + i\cfrac{\Delta t_0}{2}{\bf C}\big)^{-1}{\bf S} {\bf a}(0).
\end{eqnarray*}
Now for $\Delta t_0$ small enough, this can be simplified further because 
\begin{eqnarray*}
\big({\bf S} + i\cfrac{\Delta t_0}{2}{\bf C}\big)^{-1}  =\Big(\mathbb{I} - i\cfrac{\Delta t_0}{2}{\bf S}^{-1}{\bf C} - \cfrac{\Delta t_0^2}{4}\big({\bf S}^{-1}{\bf C}\big)^2\Big){\bf S}^{-1} + \mathcal{O}(\Delta t_0^3).
\end{eqnarray*}
Then, the time evolution operator for one time iteration has the simple form
\begin{eqnarray*}
 {\bf a}^1 = e^{-i \Delta t_0 E_0}{\bf a}(0)  + \mathcal{O}(\Delta t_0^3).
\end{eqnarray*}
From this, the discretized time evolution operator is obtained again by induction. We finally get
\begin{eqnarray*}
{\bf a}^{n+1} =  \Pi_{l=1}^{n} \Big[ \big(\mathbb{I} + i\cfrac{\Delta t_l}{2}{\bf S}^{-1}{\bf C}\big)^{-1}\big(\mathbb{I} - i\cfrac{\Delta t_l}{2}{\bf S}^{-1}{\bf C}\big)\Big]  {\bf a}(0) = e^{-i E_0\sum_{l=0}^n\Delta t_l}{\bf a}(0)  + \mathcal{O}(n\Delta t_0^3),
\end{eqnarray*}
and conclude again using similar arguments as for the Euler explicit scheme. $\Box$
\\
\\
We next state some result regarding the stability of the finite element method \eqref{Galer}.
\begin{prop}
The semi-discrete TDDE solver \eqref{Galer} with explicit Euler-based time discretization is $\ell^2$-unstable. The semi-discrete TDDE solver  with Crank-Nicolson-based time discretization, \eqref{Heun}, is $\ell^2$-stable.
\end{prop}
\noindent{\bf Proof.} From the proof of Proposition \ref{const}, stability is ensured in the explicit case when the spectral radius of the discrete evolution operator satisfies
 \begin{eqnarray*}
\rho\Big(\Pi_{l=0}^{n}  \big(\mathbb{I} -i \Delta t_l {\bf S}^{-1}\big({\bf C} + {\bf D}^{l}  \big)\big)\Big) \leq 1 \, .
\end{eqnarray*}
 In the field-free case, the spectrum was computed using the atomically balanced method. In that case, and as proven in \cite{lewin}, there is no spurious eigenvalue and all the eigenvalues are also real. We can conclude that, assuming that the eigenvalue solver is exact, the explicit Euler scheme is theoretically unstable.
\\
In the Crank-Nicolson case, the scheme reads
\begin{eqnarray*}
{\bf S}{\bf a}^{n+1} = {\bf S} {\bf a}^n -i \cfrac{\Delta t_n}{2} ({\bf C}+{\bf D}^n){\bf a}^n -i \cfrac{\Delta t_n}{2} ({\bf C}+{\bf D}^{n+1}){\bf a}^{n+1}
\end{eqnarray*}
so that, we formally have
\begin{eqnarray*}
{\bf a}^{n+1} = \Pi_{l=0}^{n}  \Big(\mathbb{I} + i\cfrac{\Delta t_l}{2}{\bf S}^{-1}\big({\bf C}+{\bf D}^{l+1}\big)\Big)^{-1}\Big(\mathbb{I} - i\cfrac{\Delta t_l}{2}{\bf S}^{-1}\big({\bf C}+{\bf D}^l\big)\Big) {\bf a}^0
\end{eqnarray*}
The requirement for stability is then that
 \begin{eqnarray*}
\rho\Big(\Pi_{l=0}^{n}  \Big(\mathbb{I} + i\cfrac{\Delta t_l}{2}{\bf S}^{-1}\big({\bf C}+{\bf D}^{l+1}\big)\Big)^{-1}\Big(\mathbb{I} - i\cfrac{\Delta t_l}{2}{\bf S}^{-1}\big({\bf C}+{\bf D}^l\big)\Big)  \Big) \leq 1
\end{eqnarray*}
We note that in the field-free case
 \begin{eqnarray*}
\rho\Big(\Pi_{l=0}^{n}  \big(\mathbb{I} + i\cfrac{\Delta t_l}{2}{\bf S}^{-1}{\bf C}\big)^{-1}\big(\mathbb{I} - i\cfrac{\Delta t_l}{2}{\bf S}^{-1}{\bf C}\big)\Big) \leq \Pi_{l=0}^{n} \rho\Big(\big(\mathbb{I} + i\cfrac{\Delta t_l}{2}{\bf S}^{-1}{\bf C}\big)^{-1}\big(\mathbb{I} - i\cfrac{\Delta t_l}{2}{\bf S}^{-1}{\bf C}\big)\Big)\leq 1 \, .
\end{eqnarray*}
Now as ${\bf S}^{-1}{\bf C}$ has real eigenvalues, this condition is trivially satisfied, and then as 
\begin{eqnarray*}
\rho\Big(\big(\mathbb{I} + i\cfrac{\Delta t_l}{2}{\bf S}^{-1}{\bf C}\big)^{-1}\big(\mathbb{I} - i\cfrac{\Delta t_l}{2}{\bf S}^{-1}{\bf C}\big)\Big)=1
\end{eqnarray*}
we have, independently on the B-spline order, $|{\bf a}^{n+1}|_2 \leq |{\bf a}^0|_2$, where $|{\bf a}^0|_2$ denote the $\ell^2$-norm of ${\bf a}^0$. \\
\\
In the laser-field case, we note that ${\bf S}^{-1}({\bf C}+{\bf D}^{n})$ does not necessarily have real eigenvalues. By regularity of the electromagnetic field, we can however deduce that ${\bf D}^{n+1} = {\bf D}^n + \mathcal{O}(\Delta t_n)$. We can reformulate the problem into
\begin{eqnarray*}
\rho \Big(\cfrac{\overline{\mathbb{I} +i \Delta t {\bf A}^n}}{\mathbb{I} + i \Delta t {\bf A}^n+ \mathcal{O}(\Delta t_n^2)}\Big) \leq 1
\end{eqnarray*}
for some complex matrix ${\bf A}^n$. The stability condition is only ensured up to a $\Delta t_n^2$ term at each time iteration. For the same reasons as described above, the following scheme is then stable:
\begin{eqnarray*}
{\bf S}{\bf a}^{n+1} = {\bf S} {\bf a}^n -i \cfrac{\Delta t_n}{2} ({\bf C}+{\bf D}^n){\bf a}^n -i \cfrac{\Delta t_n}{2} ({\bf C}+{\bf D}^{n}){\bf a}^{n+1}
\end{eqnarray*}
$\Box$
\\
We now state an important result about the convergence of \eqref{Galer} with Crank-Nicolson semi-discrete scheme in time. Although a full mathematical study of the well-posedness of 
\begin{eqnarray}\label{tdde}
i\partial_t \psi = H(t) \psi, \qquad \psi(0,\cdot) = \psi_0(\cdot)
\end{eqnarray}
would be necessary in order to determine the function space, the solution to \eqref{tdde} is living in, we can still give some relevant information about the convergence, without an explicit knowledge of these spaces.
\begin{prop}
Assume that for $\psi_0 \in H$, the solution to \eqref{eq1}, $\psi$, formally belongs to $C^1\big(0,T;V\big)$, where $V \subseteq L^2(\R^3,\C^4)$ is an Hilbert space compactly imbedded and dense in $H$ and approximated by a finite dimensional vector space $V_N$.  We also assume that $(\mathcal{B}_{j})_j :=\big([B^{(1,2)}_j,\chi_j^{(1,2)}]^T\big)_{1 \leq j \leq N}$ is a basis of $V_N$, such that $\overline{V_N}^V=V$. Then \eqref{Galer} with Crank-Nicolson-based time discretization is convergent.
\end{prop}
\noindent{\bf Sketch of the Proof.}  We follow the usual procedure, such as the one presented in \cite{7} and adapting the proof to the Dirac case. Under the above assumptions, we define the canonical projector, $P_{h_N}$, from $V$ to $V_N$ as follows
\begin{eqnarray*}
P_{h_N}\psi(t_n)  = \sum_{j=1}^N \psi_j(t_n) \otimes \mathcal{B}_{j}
\end{eqnarray*}
with $\mathcal{B}_{j} \in V_N$ and
\begin{eqnarray*}
\psi_j(t_n) = \langle \psi(t_n,\cdot) | \mathcal{B}_{j}\rangle_{L^2(\R^3,\C^4)} 
\end{eqnarray*}
The numerical approximation $\psi_{h_N}^n$ is defined as follows
\begin{eqnarray*}
\psi_{h_N}^n = \sum_{j=1}^N \psi_{h_N,j}^n\otimes \mathcal{B}_{j}
\end{eqnarray*}
where
\begin{eqnarray*}
\psi_{h_N,j}^n =  [a^{(1,2)}(t_n)_j,c^{(1,2)}(t_n)_j]^T \in \C^4
\end{eqnarray*}
and the numerical error:
\begin{eqnarray*}
\left.
\begin{array}{ccc}
e^n_{h_N} : & = &  \psi_{h_N}^n - P_{h_N}\psi(t_n)   =  \sum_{j=1}^N \Big(\psi^n_{h_N,j} - \psi_j(t_n) \Big) \otimes \mathcal{B}_{j}
\end{array}
\right.
\end{eqnarray*}
We also set:
\begin{eqnarray*}
e_j^n := \langle e_{h_N}^n | \mathcal{B}_{i}\rangle_{L^2(\R^3,\C^4)}
\end{eqnarray*}
Now from the scheme
\begin{eqnarray*}
\cfrac{1}{\Delta t_n}\langle\psi_{h_N}^{n+1}-\psi_{h_N}^n| \mathcal{B}_{j}\rangle_{L^2(\R^3,\C^4)}+\cfrac{1}{2} \langle H(t_{n+1})\psi_{h_N}^{n+1}-H(t_{n})\psi_{h_N}^n| \mathcal{B}_{j}\rangle_{L^2(\R^3,\C^4)} = 0
\end{eqnarray*}
we get
\begin{eqnarray*}
\left.
\begin{array}{ccl}
\cfrac{1}{\Delta t_n}\langle\psi_{h_N}^{n+1}-P_{h_N}\psi(t_{n+1})| \mathcal{B}_{j}\rangle_{L^2(\R^3,\C^4)} & - & \cfrac{1}{\Delta t_n}\langle\psi_{h_N}^{n}-P_{h_N}\psi(t_{n})| \mathcal{B}_{j}\rangle_{L^2(\R^3,\C^4)} \\
  +  \cfrac{1}{2} \langle H(t_{n+1})\psi_{h_N}^{n+1}-H(t_{n})\psi_{h_N}^n| \mathcal{B}_{j}\rangle_{L^2(\R^3,\C^4)} & = &   \cfrac{1}{\Delta t_n}\langle P_{h_N}\psi(t_{n}) - P_{h_N}\psi(t_{n+1}) | \mathcal{B}_{j}\rangle_{L^2(\R^3,\C^4)}
\end{array}
\right.
\end{eqnarray*}
which can also be rewritten
\begin{eqnarray*}
\left.
\begin{array}{l}
\cfrac{1}{\Delta t_n}\langle\psi_{h_N}^{n+1}-P_{h_N}\psi(t_{n+1})| \mathcal{B}_{j}\rangle_{L^2(\R^3,\C^4)} -  \cfrac{1}{\Delta t_n}\langle\psi_{h_N}^{n}-P_{h_N}\psi(t_{n})| \mathcal{B}_{j}\rangle_{L^2(\R^3,\C^4)} \\
  +  \cfrac{1}{2} \langle H(t_{n+1})\big(\psi_{h_N}^{n+1} - P_{h_N}\psi(t_{n+1})\big) | \mathcal{B}_{j}\rangle_{L^2(\R^3,\C^4)} + \cfrac{1}{2} \langle H(t_{n})\big(\psi_{h_N}^{n} - P_{h_N}\psi(t_{n})\big) | \mathcal{B}_{j}\rangle_{L^2(\R^3,\C^4)}\\
=   \cfrac{1}{\Delta t_n}\langle P_{h_N}\psi(t_{n}) - P_{h_N}\psi(t_{n+1}) | \mathcal{B}_{j}\rangle_{L^2(\R^3,\C^4)}  -   \cfrac{1}{2} \langle H(t_{n+1}) P_{h_N}\psi(t_{n+1}) + H(t_{n})P_{h_N}\psi(t_{n})| \mathcal{B}_{j}\rangle_{L^2(\R^3,\C^4)}
\end{array}
\right.
\end{eqnarray*}
and becomes
\begin{eqnarray*}
\left.
\begin{array}{l}
\cfrac{1}{\Delta t_n}\langle e_{h_N}^{n+1} - e_{h_N}^n| \mathcal{B}_{j}\rangle_{L^2(\R^3,\C^4)}   +  \cfrac{1}{2} \langle H(t_{n+1}) e_{h_N}^{n+1} | \mathcal{B}_{j}\rangle_{L^2(\R^3,\C^4)} + \cfrac{1}{2} \langle H(t_{n})e_{h_N}^n | \mathcal{B}_{j}\rangle_{L^2(\R^3,\C^4)}\\
=   \cfrac{1}{\Delta t_n}\langle P_{h_N}\psi(t_{n}) - P_{h_N}\psi(t_{n+1}) | \mathcal{B}_{j}\rangle_{L^2(\R^3,\C^4)}  -   \cfrac{1}{2} \langle H(t_{n+1}) P_{h_N}(t_{n+1}) + H(t_{n})P_{h_N}\psi(t_{n})| \mathcal{B}_{j}\rangle_{L^2(\R^3,\C^4)}
\end{array}
\right.
\end{eqnarray*}
We set
\begin{eqnarray*}
\epsilon_{h_N}^n := \cfrac{1}{\Delta t_n} \Big(P_{h_N}\psi(t_{n}) - P_{h_N}\psi(t_{n+1})\Big) -  \cfrac{1}{2} \Big(H(t_{n+1}) P_{h_N}(t_{n+1}) + H(t_{n})P_{h_N}\psi(t_{n})\Big)
\end{eqnarray*}
which is also equal to
\begin{eqnarray*}
\left.
\begin{array}{ccl}
\langle \epsilon_{h_N}^n| \mathcal{B}_{j}\rangle_{L^{2}(\R^{3},\C^{4})} & = &  \Big\langle\cfrac{1}{\Delta t_n} \Big(P_{h_N}\psi(t_{n}) - P_{h_{N}}\psi(t_{n+1})\Big) \\
 & & - \cfrac{1}{2}H(t_{n+1})\Big(P_{h_N}\psi(t_{n+1}) - \psi(t_{n+1},\cdot)\Big) - \cfrac{1}{2}H(t_{n})\Big(P_{h_N}\psi(t_{n}) - \psi(t_{n},\cdot)\Big) \\
& & +\cfrac{1}{2}H(t_{n+1})\psi(t_{n+1},\cdot) +\cfrac{1}{2} H(t_n)\psi(t_{n},\cdot)\Big| \mathcal{B}_{j}\Big\rangle_{L^2(\R^3,\C^4)}
\end{array}
\right.
\end{eqnarray*}
From
\begin{eqnarray*}
\cfrac{d}{dt}\big(P_{h_N}\psi\big) = P_{h_N}\cfrac{\partial \psi}{\partial t}
\end{eqnarray*}
and for all $j$ and all $n \geq 1$
\begin{eqnarray*}
\Big\langle \cfrac{\partial \psi}{\partial t}(t_n,\cdot)\Big| \mathcal{B}_{j}\Big\rangle_{L^2(\R^3,\C^4)} = \langle H(t_n)\psi(t_n,\cdot)| \mathcal{B}_{j}\rangle_{L^2(\R^3,\C^4)}
\end{eqnarray*}
thus
\begin{eqnarray*}
\left.
\begin{array}{ccl}
\langle \epsilon_{h_N}^n| \mathcal{B}_{j}\rangle_{L^2(\R^3,\C^4)} & = & \Big\langle \cfrac{1}{\Delta t_n} \Big(P_{h_N}\psi(t_{n}) - P_{h_N}\psi(t_{n+1})\Big)  + \cfrac{1}{2} \cfrac{\partial \psi}{\partial t}(t_n,\cdot) +\cfrac{1}{2} \cfrac{\partial \psi}{\partial t}(t_{n+1},\cdot)  \\
 & & - \cfrac{1}{2}H(t_{n+1})\Big(P_{h_N}\psi(t_{n+1}) - \psi(t_{n+1},\cdot)\Big) - \cfrac{1}{2}H(t_{n})\Big(P_{h_N}\psi(t_{n}) - \psi(t_{n},\cdot)\Big)\Big| \mathcal{B}_{j}\Big\rangle_{L^2(\R^3,\C^4)} \\
\end{array}
\right.
\end{eqnarray*}
We now set
\begin{eqnarray*}
\delta_{h_N}^n :=  \cfrac{1}{\Delta t_n} \Big(P_{h_N}\psi(t_{n}) - P_{h_N}\psi(t_{n+1})\Big)  + \cfrac{1}{2} \Big(\cfrac{\partial \psi}{\partial t}(t_n,\cdot) + \cfrac{\partial \psi}{\partial t}(t_{n+1},\cdot)\Big)
\end{eqnarray*}
and
\begin{eqnarray*}
\nu_{h_N}^n =  - \cfrac{1}{2}\Big(H(t_{n+1})\Big(P_{h_N}\psi(t_{n+1})  \psi(t_{n+1},\cdot)\Big) + H(t_{n})\Big(P_{h_N}\psi(t_{n}) - \psi(t_{n},\cdot)\Big)\Big)
\end{eqnarray*}
with $\epsilon_{h_N}^n = \delta_{h_N}^n+\nu_{h_N}^n$. Following \cite{7} and assuming that $\psi \in C^3(0,T;H)$ we get 
\begin{eqnarray*}
|\delta_{h_N}^n|_H \leq \cfrac{\Delta t_n}{8}\int_{t_n}^{t_{n+1}}\big|\cfrac{\partial^3 \psi}{\partial t^3}(s)\big|_Hds + \cfrac{1}{\Delta t_n}\int_{t_n}^{t_{n+1}}\Big|\big(I-P_{h_N}\big)\cfrac{\partial \psi}{\partial t}(s)\Big|_Hds
\end{eqnarray*}
Then, we have $\langle \nu_{h_N}^n|\mathcal{B}_{i}\rangle_{L^2(\R^3,\C^4)}$ that goes to zero when $h \rightarrow 0$, due to the density of $V_N$ is $V$. 
\\
Now, from
\begin{eqnarray*}
\left.
\begin{array}{l}
\cfrac{1}{\Delta t_n}\langle e_{h_N}^{n+1} - e_{h_N}^n| \mathcal{B}_{j}\rangle_{L^2(\R^3,\C^4)}   +  \cfrac{1}{2} \langle H(t_{n+1}) e_{h_N}^{n+1}  + H(t_{n})e_{h_N}^n | \mathcal{B}_{j}\rangle_{L^2(\R^3,\C^4)} = \Delta t_n \langle \delta_{h_N}^n + \nu_{h_N}^n| \mathcal{B}_{j}\rangle_{L^2(\R^3,\C^4)}
\end{array}
\right.
\end{eqnarray*}
we have, without approximation
\begin{eqnarray*}
{\bf S}{\bf e}^{n+1}= {\bf S}{\bf e}^n -i \cfrac{\Delta t_n}{2} \big({\bf C} + {\bf D}^{n} \big){\bf e}^n - i \cfrac{\Delta t_n}{2} \big({\bf C} + {\bf D}^{n+1} \big){\bf e}^n + \Delta t_n({\bf \delta}^n + {\bf \nu}^n)
\end{eqnarray*}
where 
\begin{eqnarray*}
\left.
\begin{array}{ccl}
{\bf e}^n  & = & [ \psi^{(1)}_1(t_n)-a^{(1),n}_{1},\cdots,\psi^{(1)}_{h_N}(t_n)-a^{(1),n}_{N},\psi^{(2)}_1(t_n)-a^{(2),n}_{1},\cdots,\psi^{(2)}_{h_N}(t_n)-a^{(2),n}_{N},\\
& & \psi^{(3)}_1(t_n)-c^{(1),n}_{1},\cdots,\psi^{(3)}_{h_N}(t_n)-c^{(1),n}_{N},\psi^{(4)}_1(t_n)-c^{(2),n}_{1},\cdots,\psi^{(4)}_{h_N}(t_n)-c^{(2),n}_{N}]
\end{array}
\right.
\end{eqnarray*}
and ${\bf \delta}^n = \big(\langle \delta_{h_N}^n| \mathcal{B}_{j} \rangle_{L^2(\R^3,\C^4)}\big)_j$, ${\bf \nu}^n = \big(\langle \nu_{h_N}^n| \mathcal{B}_{j} \rangle_{L^2(\R^3,\C^4)}\big)_j$. Now we deduce
\begin{eqnarray*}
\Big({\bf S} + i \cfrac{\Delta t_n}{2} \big({\bf C} + {\bf D}^{n+1} \big)\Big){\bf e}^{n+1}= \Big({\bf S} -i \cfrac{\Delta t_n}{2} \big({\bf C} + {\bf D}^{n} \big)\Big){\bf e}^n + \Delta t_n({\bf \delta}^n+{\bf \nu}^n)
\end{eqnarray*}
Then 
\begin{eqnarray*}
{\bf e}^{n+1}= \Big({\bf S} + i \cfrac{\Delta t_l}{2} \big({\bf C} + {\bf D}^{l+1} \big)\Big)^{-1}\Big({\bf S} -i \cfrac{\Delta t_l}{2} \big({\bf C} + {\bf D}^{l} \big)\Big){\bf e}^n + \Delta t_n\Big({\bf S} + i \cfrac{\Delta t_l}{2} \big({\bf C} + {\bf D}^{n+1} \big)\Big)^{-1}({\bf \delta}^n+{\bf \nu}^n)
\end{eqnarray*}
Finally from
\begin{eqnarray*}
|\psi_{h_N}^n - \psi(t_n,\cdot)|_H \leq |\psi_{h_N}^n - P_{h_N}\psi(t_n)|_H + |(I -P_{h_N}) \psi(t_n,\cdot)|_H = |e_{h_N}^n|_H + |(I -P_{h_N}) \psi(t_n,\cdot)|_H
\end{eqnarray*}
we formally conclude of the convergence of the method, as in \cite{7}. We note again that this conclusion is only valid under strong reasonable assumptions. $\Box$

\section{Numerical implementation}\label{sec:impl}

The numerical method described in previous sections have been implemented in a high performance parallel  code. This is required because the calculation of physical observables entails a large amount of computational resources: the typical time step should obey $\delta t \lesssim 1/mc^{2}$ to guarantee high precision \cite{FillionGourdeau20121403} while the dynamics of typical external laser fields occurs on much larger time scales. Moreover, for QED calculations, every negative energy states has to be evolved in time and thus, demand a large number of time evolution calculations. 

\begin{figure}[!ht]
\begin{center}
\hspace*{1mm}\includegraphics[height=7cm, keepaspectratio]{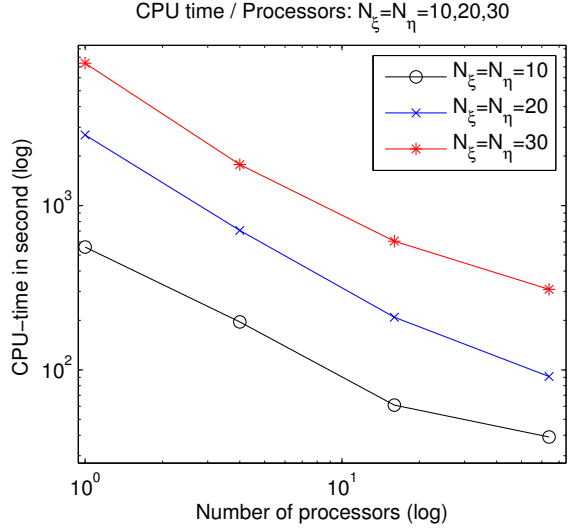}
\caption{CPU-time $/$ processors for time evolution of $\mbox{Th}_2^{179+}$ molecule ($10^4$ iterations) for $N_{\xi}=N_{\eta}=10$ and $N_{\xi}=N_{\eta}=20$ and $N_{\xi}=N_{\eta}=30$}
\label{CPU}
\end{center}
\end{figure}

The parallelization is performed by using the capabilities of the PETSc \cite{petsc-efficient} and SLEPc \cite{Hernandez:2005:SSF} linear algebra libraries. Because the B-spline basis functions have compact but overlapping support, it is not convenient to employ a standard domain decomposition, as in \cite{FillionGourdeau20121403}, for example. Rather, the parallelization is accomplished by distributing the solution vector $\mathbf{a}$ on many processors as
\begin{eqnarray}
\mathbf{a}(t)|_{\mathrm{proc\;1}} &=& [ a^{(1)}_{1}(t),a^{(2)}_{1}(t),c^{(1)}_{1}(t),c^{(2)}_{1}(t), \cdots, a^{(1)}_{n}(t),a^{(2)}_{n}(t),c^{(1)}_{n}(t),c^{(2)}_{n}(t)] \nonumber \\
\mathbf{a}(t)|_{\mathrm{proc\;2}} &=& [ a^{(1)}_{n+1}(t),a^{(2)}_{n+1}(t),c^{(1)}_{n+1}(t),c^{(2)}_{n+1}(t), \cdots, a^{(1)}_{2n}(t),a^{(2)}_{2n}(t),c^{(1)}_{2n}(t),c^{(2)}_{2n}(t)] \nonumber \\
\vdots &=& \vdots \nonumber \\
\mathbf{a}(t)|_{\mathrm{proc}\;M} &=& [ a^{(1)}_{N-n+1}(t),a^{(2)}_{N-n+1}(t),c^{(1)}_{N-n+1}(t),c^{(2)}_{N-n+1}(t), \cdots, a^{(1)}_{N}(t),a^{(2)}_{N}(t),c^{(1)}_{N}(t),c^{(2)}_{N}(t)].
\end{eqnarray} 
Here, $N$ is the number of basis functions, $M$ is the number of processors and $n$ is the number of basis function stored on one processor: they are related by $N=M \cdot n$. This ordering of coefficients insures that all the spinor component contributions with the same support are stored on the same processor. Moreover, it is consistent with the PETSc parallel matrix storage, which adopts a row-wise storage type. Then, the number of entries for the matrices $\mathbf{S}$, $\mathbf{C}$ and $\mathbf{D}$ is the same on all processors, insuring an equal load on every processor. The calculation of these matrices does not require any inter-processor communications and thus, this part of the calculation is embarrassingly parallel. Communications are required when the linear system or the general eigenvalue problems are solved: these operations are dealt with efficiently by the PETSc and SLEPc libraries. These features make for a very efficient parallel code with an excellent parallel speedup.   \\
In order to show the efficiency of the proposed parallelization, we study the time evolution of the wavefunction for a dithorium two-center system subjects to an external electric field. Data are respectively as follows: $N_{\xi}=N_{\eta}=10$, $N_{\xi}=N_{\eta}=20$, then $N_{\xi}=N_{\eta}=30$, with B-spline order fixed to $5$. The time step is fixed to $10^{-6}$ and $10^4$ time iterations are performed. We report in logscale in  Fig. \ref{CPU}, the computational time using respectively 1, 4, 16, 32 and 64 processors. 
Notice, that the discrepancy in the scalability graph, observed for 64 processors, is a simple consequence of the moderate size of this benchmark. 
\section{Numerical results}\label{NUM}
\begin{figure}[!ht]
\begin{center}
\hspace*{1mm}\includegraphics[height=7cm, keepaspectratio]{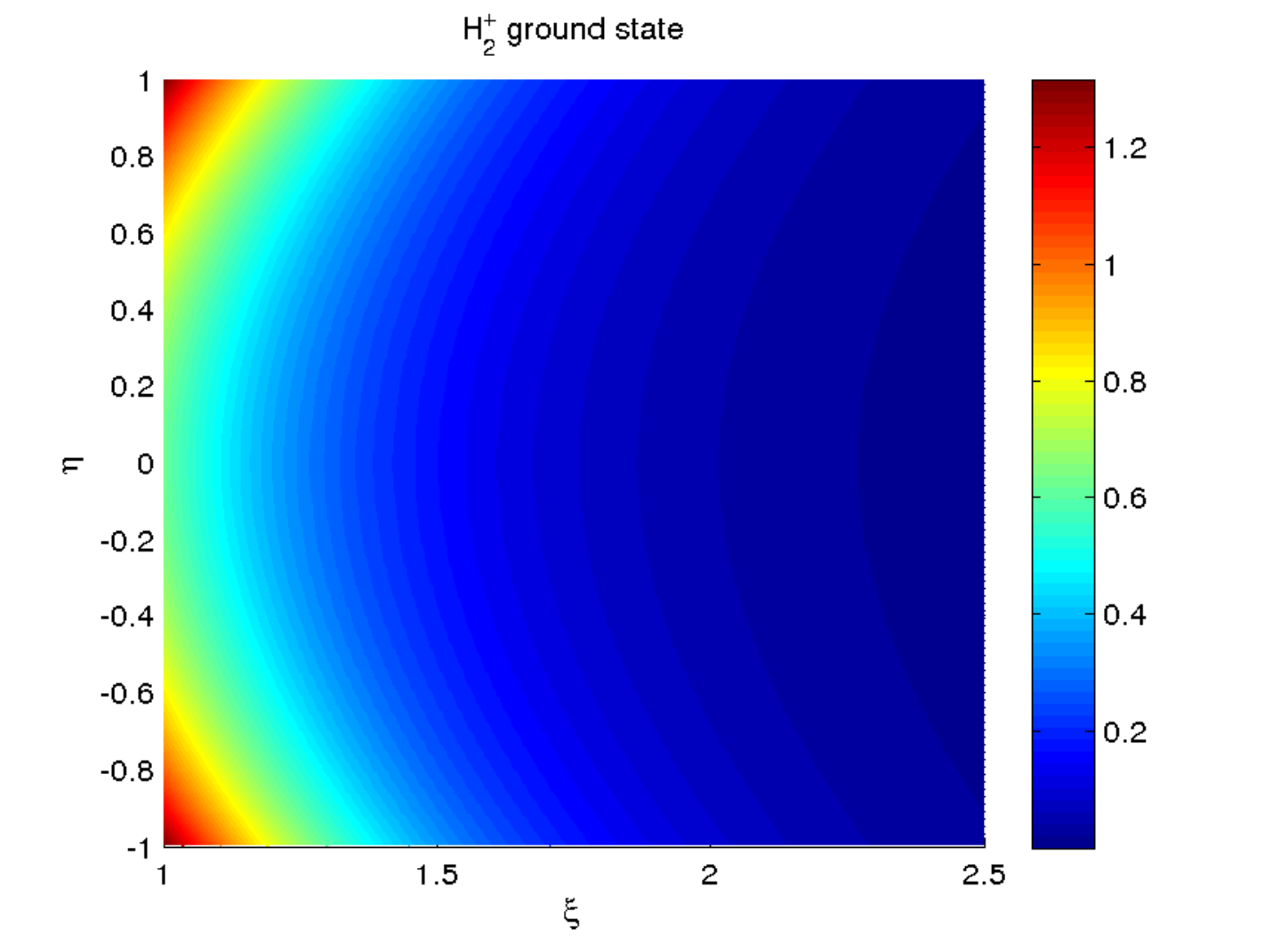}
\caption{$\mbox{H}_2^+$ ground state, represented on $1024\times 1024$ grid points}
\label{h2plus}
\end{center}
\end{figure}

\begin{figure}[!ht]
\begin{center}
\hspace*{1mm}\includegraphics[height=7cm, keepaspectratio]{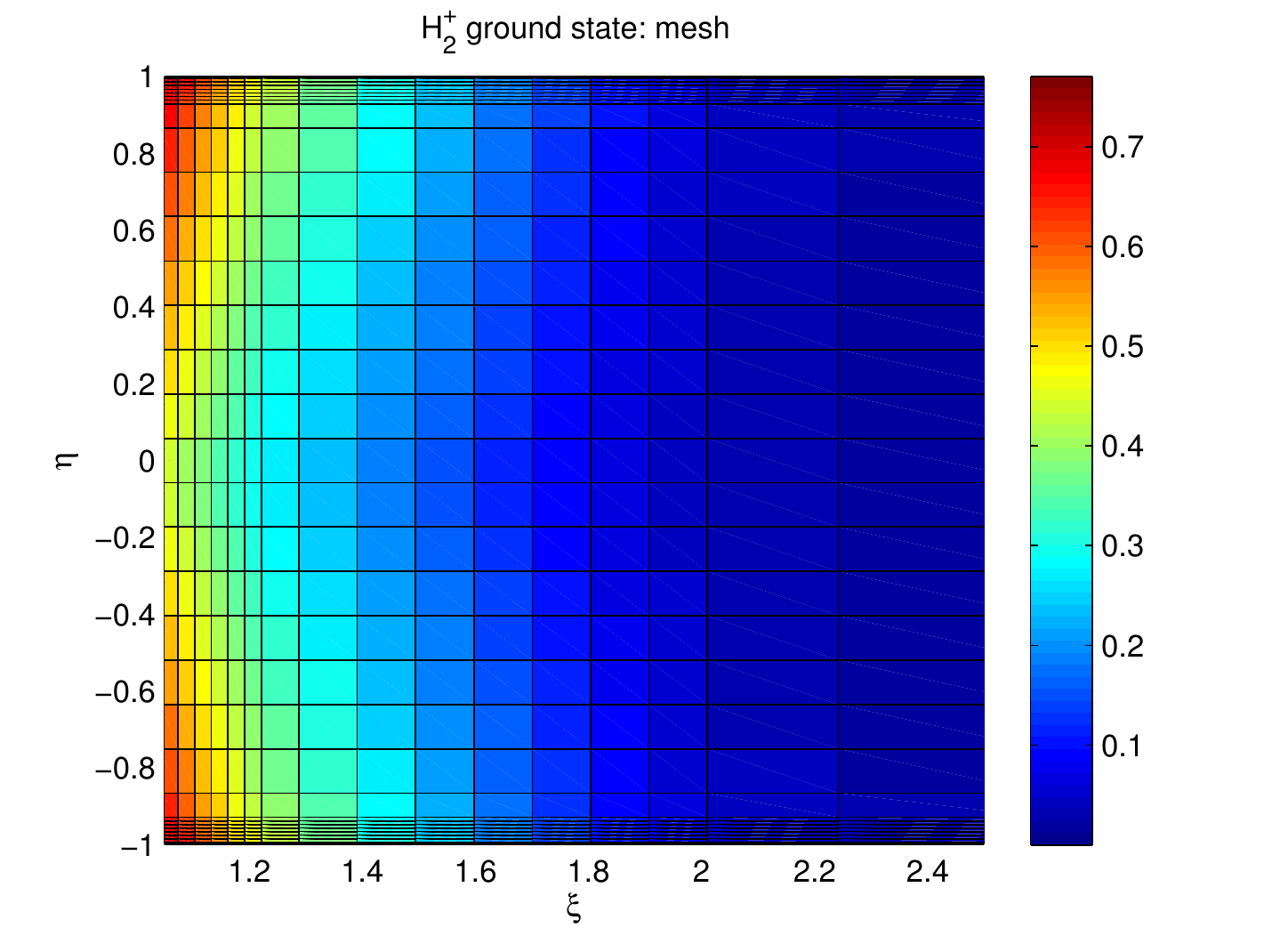}
\caption{Mesh $32\times32$ grid points in prolate spheroidale coordinates, and corresponding $\mbox{H}_2^+$ ground state computed with $20$ basis functions}
\label{grid}
\end{center}
\end{figure}

This section is devoted to the numerical validation of the B-spline method presented in Sections \ref{TIDE}, \ref{TDDE}. Detailed physical properties of system under consideration, will be studied in a forthcoming paper, specifically dedicated to quantum relativistic particles subject to external classical field. In this paper, the considered particles are dihydrogen ($Z_{1,2}=1$) or dithorium ($Z_{1,2}=90$). The angular momentum is fixed to $j_z=1/2$. 
From the numerical point of view, several parameters have to be fixed. We recall that $N_{\xi,\eta}$ denote the number of elements in each coordinates, and $N^*$ the total number of basis functions. In Fig. \ref{h2plus}, we illustrate the $\mbox{H}_2^+$ ground state (with $R=1$) in the prolate spheroidal coordinates, with $N_{\xi}=N_{\eta}=4$ and only $N^*=20$ basis functions. Note in particular, the positions of the nuclei, at the left corners of the grid.  
Fig. \ref{grid} reports the grid structure for $32\times 32$ grid, and the corresponding numerical $\mbox{H}_2^+$ ground state.
\subsection{Convergence for TIDE}
In this section, we investigate the numerical convergence of the atomic$\&$kinetic balance technique with a B-spline basis.  The tests are similar to those presented in \cite{FFG}. More specifically, we study and calculate the ground state of $\mbox{Th}_{2}^{179+}$ (dithorium) for which $Z_{1,2}=90$, and $\mbox{H}_{2}^{+}$ (dihydrogen) for which $Z_{1,2}=1$. The semi inter atomic distance is set to $R = \frac{1}{90} \approx 0.011111$ a.u. for $\mbox{Th}_{2}^{179+}$ and to $R \approx 1.0$ a.u. for $\mbox{H}_{2}^{+}$,  while the angular momentum is taken as $j_{z} = 1/2$. The results for the calculation of the ground state binding energy using B-splines of order 7 and different mesh sizes are shown in Table \ref{table:res_conv_H2} and \ref{table:res_conv_Th179} for $\mbox{H}_{2}^{+}$, and $\mbox{Th}_{2}^{179+}$ respectively.

The results presented in this table show the convergence of the method as the number of elements, at fixed B-spline order, are increased. The results obtained are very accurate, although there is a small difference ($\approx 10^{-8}$\% and $\approx 10^{-4}$\% for $\mbox{H}_{2}^{+}$ and $\mbox{Th}_{2}^{179+}$, respectively) between our results and the results presented in \cite{Kullie2004215}. This difference can be explained by a different choice of boundary conditions, different element formulation and different treatment of the Coulomb singularity.   The B-spline basis functions, being polynomial with integer powers, are unable to reproduce exactly this feature. Moreover, we have that
\begin{eqnarray}
 \gamma_{\rm H} \approx 0.999947 \;\; \mbox{and} \;\; \psi \sim r_{1,2}^{-0.000053} \\
\gamma_{\rm Th} \approx 0.568664 \;\; \mbox{and} \;\; \psi \sim r_{1,2}^{-0.431336}
\end{eqnarray}
where $\gamma_{\rm H,Th}$ are the gamma associated with a hydrogen or thorium atom. It is clear from this that the behavior of the wavefunction is much closer to a power law for dihydrogen and therefore, is better reproduced by the B-splines and thus, has a faster convergence. 

One possible cure to this is to use another prefactor in the basis function that mimics the correct behavior. For instance, it has been proposed to multiply the basis functions in (\ref{eq:basis_func_def}) by \cite{PhysRevA.48.2700,Kullie1999307,Kullie2004215} by
\begin{eqnarray}
G'(\xi,\eta) = r_{1}^{-1+\gamma_{1}}  r_{2}^{-1+\gamma_{2}} 
\end{eqnarray}
with
\begin{eqnarray}
 r_{1} = (\xi+\eta)R, \; r_{2} = (\xi-\eta)R .
\end{eqnarray}
where
\begin{eqnarray}
\gamma_{1,2} = \sqrt{\left(|j_{z}| + \frac{1}{2}\right)^{2} - \alpha^{2} Z_{1,2}^{2}} .
\end{eqnarray}
and $r_{1,2}$ are the internuclear distances between 1 and 2. In ground state calculations, we have $j_{z} = 1/2$ and thus, $0 < \gamma_{1,2} < 1$ for $Z_{1,2}<137$. Therefore, the wavefunction has a non-integer power-law behavior close to the singularity at $r=0$.

The main issue with this method is that the derivative in the functionals become singular. To cope with this, a singular coordinate transformation can be performed that allows to transform the singular non-integer behavior near the nuclei to a polynomial approximation \cite{PhysRevA.48.2700,Kullie1999307}.

\begin{table}[h]
\caption{Results of the numerical computation for the ground state of $\mbox{H}_{2}^{+}$ for different mesh sizes and B-spline of order 7. Here, $N_{\xi,\eta}$ are the number of elements in each coordinates while $N^{*}$ is the total number of basis functions utilized. The maximum coordinate was fixed to $\xi_{\rm max}=30$ a.u. and the angular momentum to $j_{z} = 1/2$. The calculations are to be compared with the results from \cite{Kullie2004215} where the authors obtained E$_{\mathrm{H}_{2}^{+}}$ = -1.10264158103 a.u..}
\centering
\begin{tabular}{lllccc}
\hline
$N_{\xi}$ & $N_{\eta}$ & $N^{*}$  &  \multicolumn{3}{c}{E$_{\mathrm{H}_{2}^{+}}$ (a.u.)} \\
	  &	       &          & Min-max & Kinetic & Atomic\\
\hline
8 & 8 & 182 & -1.102590816884 & -1.102590816895 &   -1.102590816899\\
10 & 10 & 240 & -1.102638533873 & -1.102638533934 & -1.102638533914\\
12 & 12 & 306 & -1.102641366239 & -1.102641366228 & -1.102641366222\\
14 & 14 & 380 & -1.102641554428 & -1.102641554501 & -1.102641554498\\
16 & 16 & 462 & -1.102641577089 & -1.102641577085 & -1.102641577079\\
18 & 18 & 552 & -1.102641580210 & -1.102641580229 & -1.102641580219\\
20 & 20 & 650 & -1.102641580782 & -1.102641580825 & -1.102641580823\\
\hline
\end{tabular} 
\label{table:res_conv_H2}
\end{table}

\begin{figure}[!ht]
\begin{center}
\hspace*{1mm}\includegraphics[height=6cm, keepaspectratio]{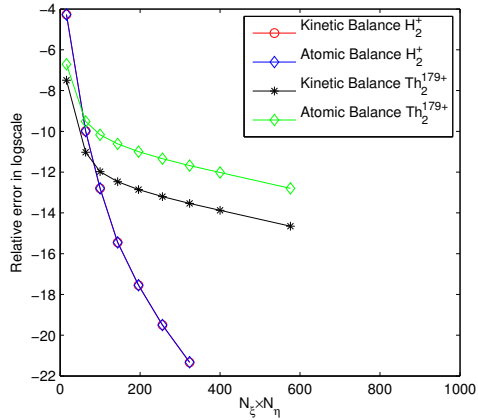}
\caption{Logarithm of relative error of the groundstate energy as function of $N_{\xi}\times N_{\eta}$ (total number of nodes) at B-spline order $7$, for $\mbox{Th}_2^{179+}$ and $\mbox{H}_2^+$}
\label{vp_conv}
\end{center}
\end{figure}

\begin{table}[h]
\caption{Results of the numerical computation for the ground state of $\mbox{Th}_{2}^{179+}$ for different mesh sizes and B-spline of order 7. Here, $N_{\xi,\eta}$ are the number of elements in each coordinates while $N^{*}$ is the total number of basis functions utilized. The maximum coordinate was fixed to $\xi_{\rm max}=15$ a.u. and the angular momentum to $j_{z} = 1/2$. The calculations are to be compared with the results from \cite{Kullie2004215} and \cite{0953-4075-43-23-235207} where the authors obtained E$_{\mathrm{Th}_{179}^{+}}$ = -9504.756746922 a.u. and E$_{\mathrm{Th}_{179}^{+}}$ = -9504.752 a.u..}
\centering
\begin{tabular}{lllccc}
\hline
$N_{\xi}$ & $N_{\eta}$ & $N^{*}$  &  \multicolumn{3}{c}{E$_{\mathrm{Th}_{2}^{179+}}$  (a.u.)} \\
	  &	       &          & Min-max & Kinetic & Atomic \\
\hline
8  & 8  & 182 & -9503.998584802 & -9504.592903093489 & -9503.999825720\\
10 & 10 & 240 & -9504.333585765 & -9504.687718599949 & -9504.333923392\\
12 & 12 & 306 & -9504.466070634 & -9504.711184750768 & -9504.466246166\\
14 & 14 & 380 & -9504.539502492 & -9504.722872750701 & -9504.539637808\\
16 & 16 & 462 & -9504.586247153 & -9504.730120406488 & -9504.586369144\\
18 & 18 & 552 & -9504.618392312 & -9504.735095027911 & -9504.618508491\\
20 & 20 & 650 & -9504.641636959 & -9504.738703758736 & -9504.641750168\\
24 & 24 & 870 & -9504.672557123 & -9504.743524797539 & -9504.672667124\\
30 & 30 & 1260& -9504.698874401 & -9504.747650405050 & -9504.698989287\\
\hline
\end{tabular} 
\label{table:res_conv_Th179}
\end{table}
From Table \ref{table:res_conv_Th179}, we can deduce the rate of convergence to the groundstate energy of reference (computed with $N_{\xi}=N_{\eta}=30$). We represent in Fig. \ref{vp_conv}, the logarithm of the relative error as a function $N_{\xi}\times N_{\eta}$ in both cases.  This graph also illustrates, that the strength of the singularity is responsible for a deterioration of the overall convergence rate of the method for large $Z$.
We observe in the {\it atomic balance} case, that the overall convergence behavior for $\mbox{Th}_{2}^{179+}$ is quite similar to the min-max approach and is not as good as in the {\it kinetic balance} case. Although, we do not have a clear explanation for that, we think that the discrepancy in the convergence, is due to the presence of $V_c$, with $Z_{1,2}$ large (for small $Z_{1,2}$, see Table \ref{table:res_conv_H2}, the convergence rate is roughly similar to the kinetic case), in the variational intregals defining $\mathbf{C}$, $\mathbf{D}$, $\mathbf{S}$. It is then challenging to numerically maintain an high order of accuracy close to the potential singularities. This will be subject to future investigation. 
\subsection{Convergence with increasing B-spline order}
\begin{figure}[!ht]
\begin{center}
\hspace*{1mm}\includegraphics[height=6cm, keepaspectratio]{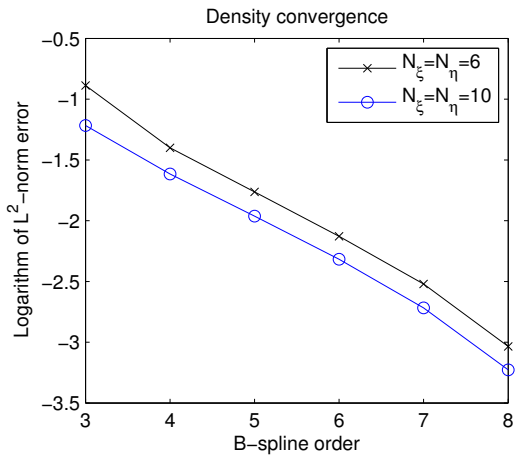}
\caption{Logarithm of $L^2$-norm error on total density as function of B-spline order (kinetic balance operator): $N_{\xi}=N_{\eta}=10$ and $N_{\xi}=N_{\eta}=6$}
\label{ABvsKB}
\end{center}
\end{figure}

 In the following test, we compute the following error on the total density, for $\mbox{Th}_2^{179+}$:
\begin{eqnarray*}
e^{(p)}:=\big\|\rho_{g}-\rho^{(p)}\big\|_{L^2(\R^3,\R)}
\end{eqnarray*}
for different orders $p$, where $\rho^{(p)}$ denotes the numerical density constructed from an order $p$, B-spline function basis, with $N_{\xi,\eta}$ grid points in directions $\xi,\eta$, and $N^*$ basis functions, and $\rho_g$ is a solution of reference constructed with very high order B-splines.  In other words, we compute for fixed mesh, the error as a function of the B-spline order. This unusual way to show the convergence of the method is justified by the non-nestedness of meshes for $\big\{N_{\xi}/2^i$, $N_{\eta}/2^i$, $i\geq 1\}$ in prolate spheroidal coordinates, making it hard (without using very high order interpolation methods), to determine numerically the order of the overall scheme. In addition, the boundary conditions, the singularity, the number of knots, the integration method and its order, have all an effect on the overall order of the scheme. We here then show that the higher the B-spline order, the smaller the relative error, justifying the use of high order B-splines.  For $\mbox{H}_2^+$, we report in Fig. \ref{ABvsKB} the semilogscale of the $L^2$-error of the overall density $\rho(t,\xi,\eta)=\big(\sum_{i=1}^4|\psi_i(t,\xi,\eta)|^2)^{1/2}$ for different B-spline order, and for $N_{\xi}=N_{\eta}=6$ and $N_{\xi}=N_{\eta}=10$. Results are shown using the {\it kinetic balance operator}. These tests show that as expected, the $L^2-$norm error (with respect to a a solution of reference) is function of the power of the order of the B-spline (semilogscale in ordinates is used).  Note that in all the computations, the number of Gauss-Legendre points has been fixed to $64$. 
\subsection{Energy spectra of diatomic molecules}
Energy spectra are calculated using a mesh of 30$\times$30 elements. The other parameters are set to the same values as in the last section where the convergence of the ground state was discussed. The value of the binding energies in the mass gap ($[-mc^{2},mc^{2}]$) which corresponds to bound states are shifted by $mc^{2}$ to have a comparison with non-relativistic results. The values in the continua however are not shifted and calculated with the Rayleigh-Ritz method only. The results of the dithorium spectrum can be compared to the ones in \cite{0953-4075-42-5-055002}. Both are generally in good agreement, although a small discrepancy can be seen for the higher excited states.

In the Rayleigh-Ritz method, the $n_{\rm binding}$ bound state energies shown in Tables \ref{table:res_spec_H2} and \ref{table:res_spec_Th2} correspond to the $2N+1$ to $2N+1+n_{\rm binding}$ eigenvalues of the matrix $\mathbf{C}$ (once the eigenvalues are in increasing order). The other eigenvalues can be associated to the ``discretized'' negative (the first to the $2N$'th eigenvalues) and positive (the $2N+2+n_{\rm binding}$'th to the $4N$'th eigenvalues) energy continua. 

The convergence of the excited states is very similar to the ground state: all the values are approached from above and the order of convergence is close to the one of the ground state. The same is true for the states in the positive energy continuum, that is for $E \geq mc^{2}$. For the negative energy states, the convergence occurs from below, but otherwise, follows the same trends as the other cases. The energy values in the continuua (especially their smallest and largest eigenvalues) depend on the size of the domain. In the dithorium calculation, the domain was smaller which yielded less accurate value in the continuua (not shown in the table) but better accuracy of the bound states. In all cases, the eigenvalues of the positive and negative energy continua accumulate at the points $mc^{2}$ and $-mc^{2}$, respectively.

\begin{table}[h]
\caption{Results of the numerical computation for the spectrum of $\mbox{H}_{2}^{+}$ for a mesh size of 30$\times$30 and B-spline of order 7. The states of the positive and negative continua are computed with the Rayleigh-Ritz, Min-Max and Atomic Balance methods. The first 5 states are shown.}
\centering
\begin{tabular}{llll|lll}
Bound &  \multicolumn{3}{c}{Binding energy  (a.u.)} & & Negative & Positive\\
states& Min-max & RR & Atomic &  &continuum (a.u.) & continuum (a.u.)\\
1 & -1.1026413662& -1.1026415808& -1.1026415808 & 1& -18778.95240& 18778.86549\\
2 & -0.6675525594& -0.6675527718& -0.6675527718 & 2& -18778.95792& 18778.86561\\
3 & -0.4287795568& -0.4287811584& -0.4287810919 & 3& -18778.96471& 18778.86562\\
4 & -0.3608697621& -0.3608710695& -0.3608690590 & 4& -18778.97284& 18778.86741\\
5 & -0.2554175614& -0.2554197033& -0.2553343110 & 5& -18778.98233& 18778.86746\\
\end{tabular} 
\label{table:res_spec_H2}
\end{table}

%
%

\begin{table}[h]
\caption{Results of the numerical computation for the spectrum of $\mbox{Th}_{2}^{179+}$. The mesh size is indicated on the second line. The B-splines are of order 7.}
\centering
\begin{tabular}{llllll}
States &  \multicolumn{2}{l}{Naive RR} &  RR & Min-max & Atomic\\
	& $14\times14$ & $30 \times 30$ & $30 \times 30$ & $16 \times 16$ & $30 \times 30$ \\
1 & -9504.6525442 & -9504.7243225 & -9504.7475523 & -9504.5862992 & -9504.6416456\\ 
2 & -6815.3652913 & -6815.4657298 & -6815.5599111 & -6815.3230307 & -6815.3865298\\ 
3 & -4127.8799531 & -4127.8877478 & -4128.1451137 & -4127.8197047 & -4127.8457787\\ 
4 & -3374.4958326 & -3374.5117016 & -3374.5143753 & -3374.4569981 & -3374.4767336\\ 
5 & -2564.1326367 & -2564.1559253 & -2564.1719708 & -2564.0744037 & -2564.0918230\\ 
6 & -2455.9453341 & -2455.9537953 & -2455.9600280 & -2455.8837393 & -2455.9016668\\ 
7 & -2010.6579407 & -2010.6535604 & -2010.4321103 & -2010.4241948 & -2010.4261981\\ 
8 & -1918.5275474 & -1918.4056980 & -1915.7178408 & -1915.6761267 & -1915.6853488 \\ 
9 & -1649.5111100 & -1649.2929148 & -1643.9543595 &  -1643.9320665 & -1643.9395109\\ 
10 & -1349.5529034 & -1344.0855870 & -1313.8071916 & -1313.7606899 & -1313.7699129\\ 
11 & -1339.1123032 & -1333.5368147 & -1303.6850950 & -1303.6580541 & -1303.6660492\\ 
spurious & -1218.2113620 & -1204.6990945 &   & \\ 
12 & -1169.3956263 & -1159.1761393 & -1089.6415827 & -1089.6356220 & -1089.6370783\\ 
13 & -1138.5709512 & -1131.0151665 & -1084.3699127 & -1084.3519981 & -1084.3522895\\ 
14 & -1046.2053120 & -1045.4764538 & -1028.1920826 & -1028.1912423 & -1028.1920249\\ 
15 & -1018.4013912 & -984.5252901 & -969.6816867 & -969.64172165 & -969.6482618\\ 
\end{tabular} 
\label{table:res_spec_Th2}
\end{table}
Notice that although, theoretically (see Theo \ref{theoKB}), spurious states can be generated using a kinetic balance operator, in the numerical tests we performed with this balance, only physical states were generated, while using the kinetic balance.

\subsection{Numerical tests for TDDE}
\begin{figure}[!ht]
\begin{center}
\hspace*{1mm}\includegraphics[height=4cm, keepaspectratio]{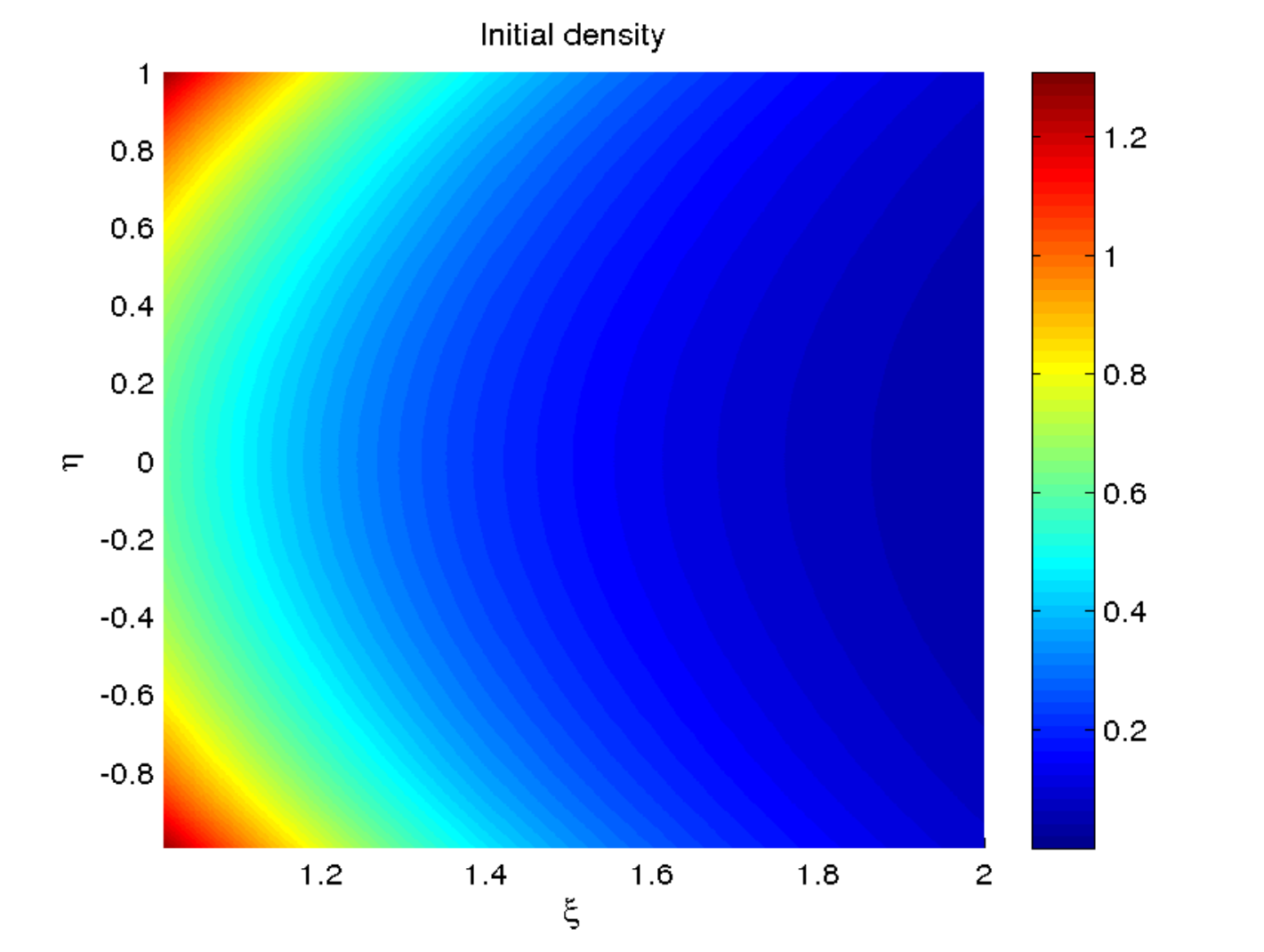}
\hspace*{1mm}\includegraphics[height=4cm, keepaspectratio]{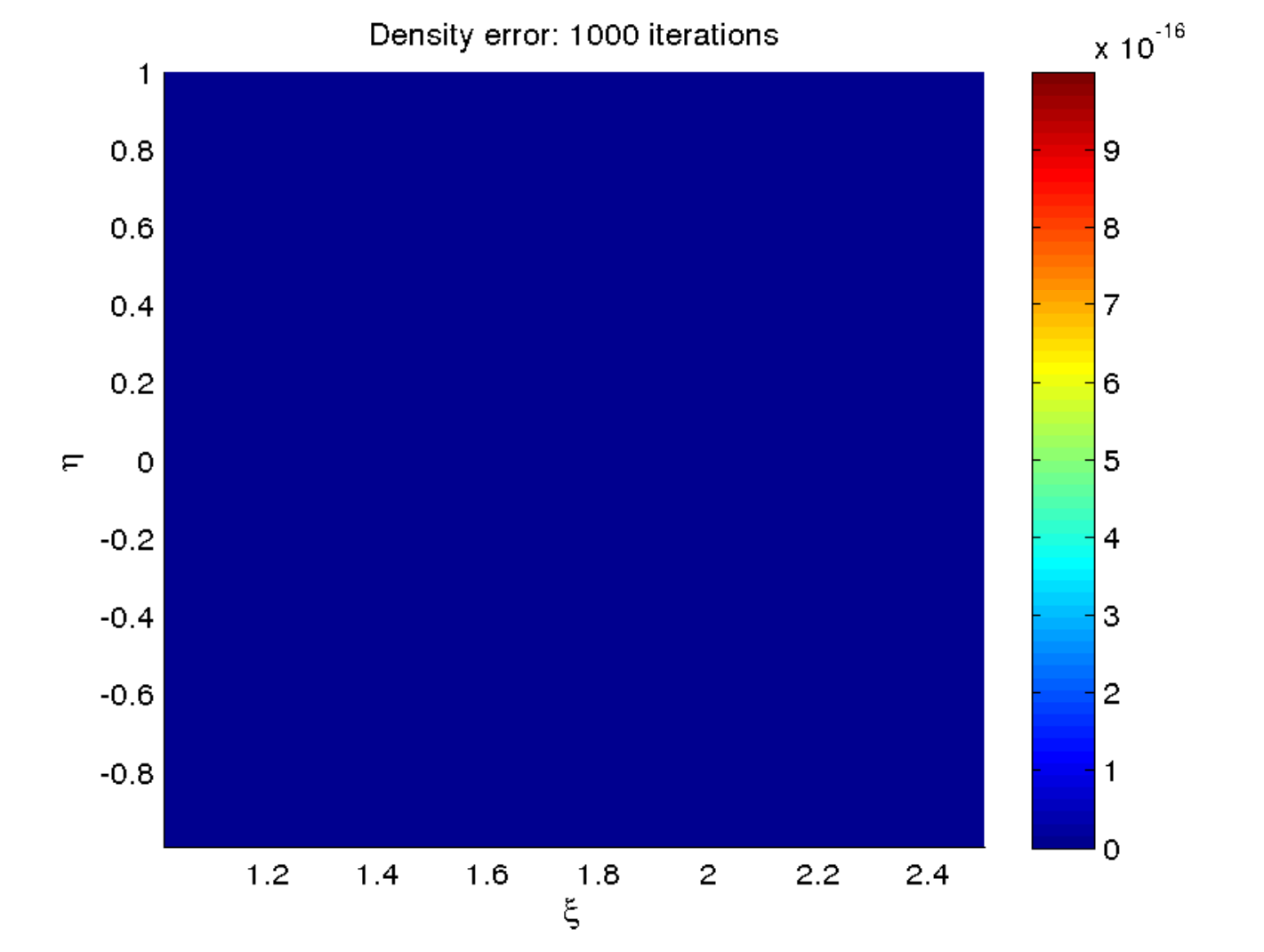}
\hspace*{1mm}\includegraphics[height=4cm, keepaspectratio]{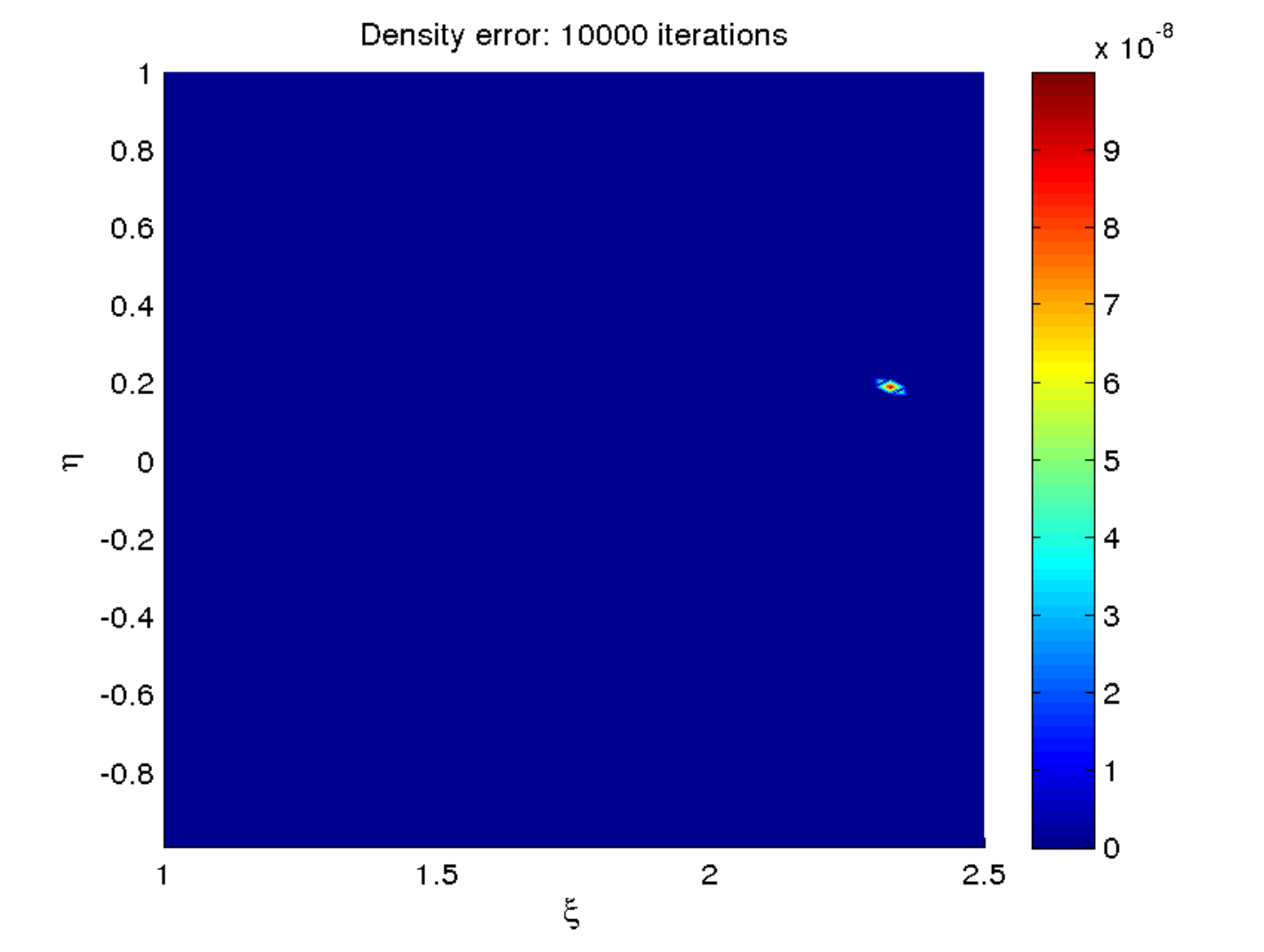}
\caption{Density comparison: $|\rho(t_f,\cdot)-\rho_0(\cdot)|$ after $10^3$ and $10^4$ iterations computed with $N^*=182$ basis functions and order 7 B-spline.  Representation on $256\times256$ grid points}
\label{nofield}
\end{center}
\end{figure}

\begin{figure}[!ht]
\begin{center} 
\hspace*{1mm}\includegraphics[height=2.8cm, keepaspectratio]{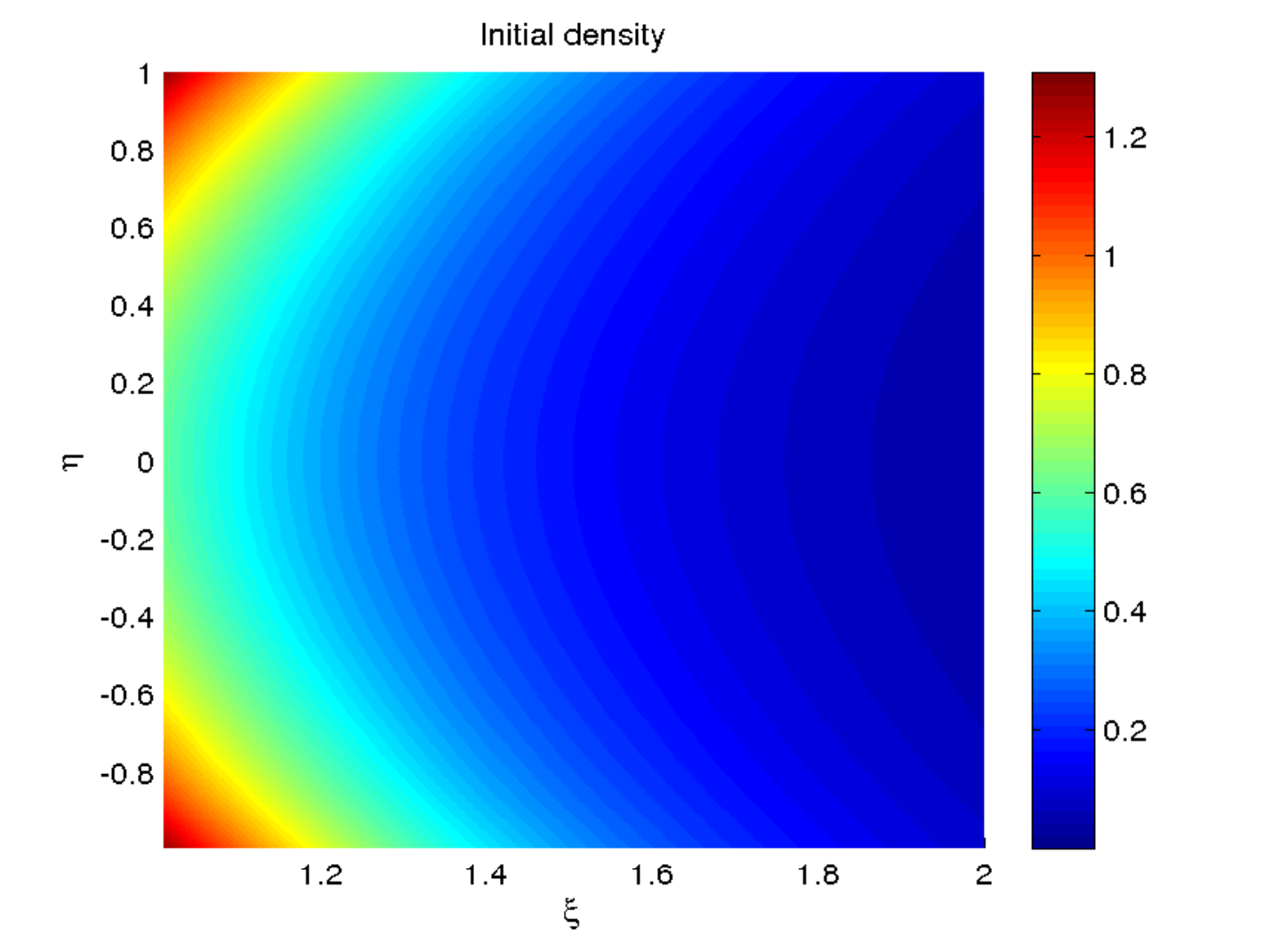}
\hspace*{1mm}\includegraphics[height=2.8cm, keepaspectratio]{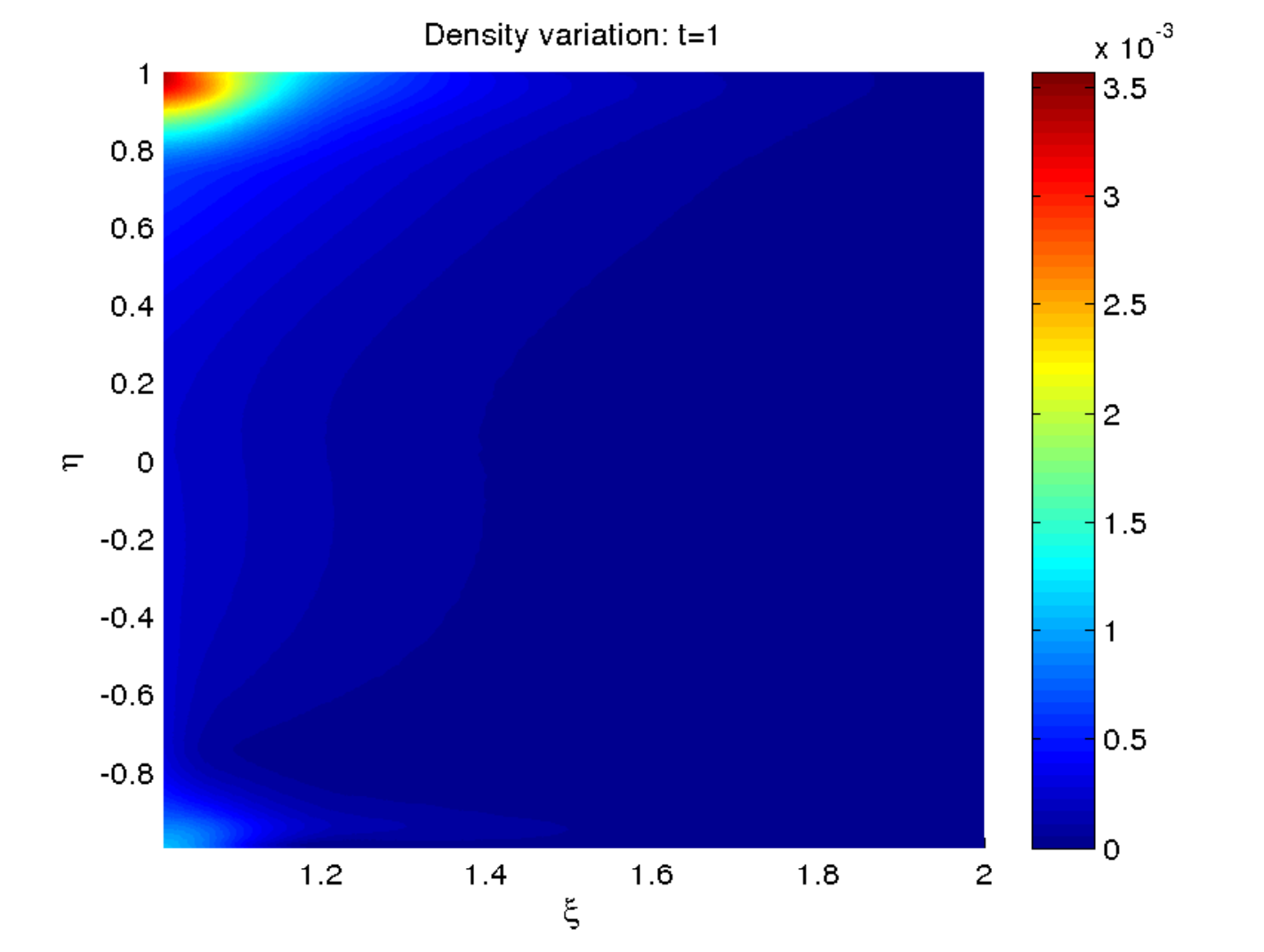}
\hspace*{1mm}\includegraphics[height=2.8cm, keepaspectratio]{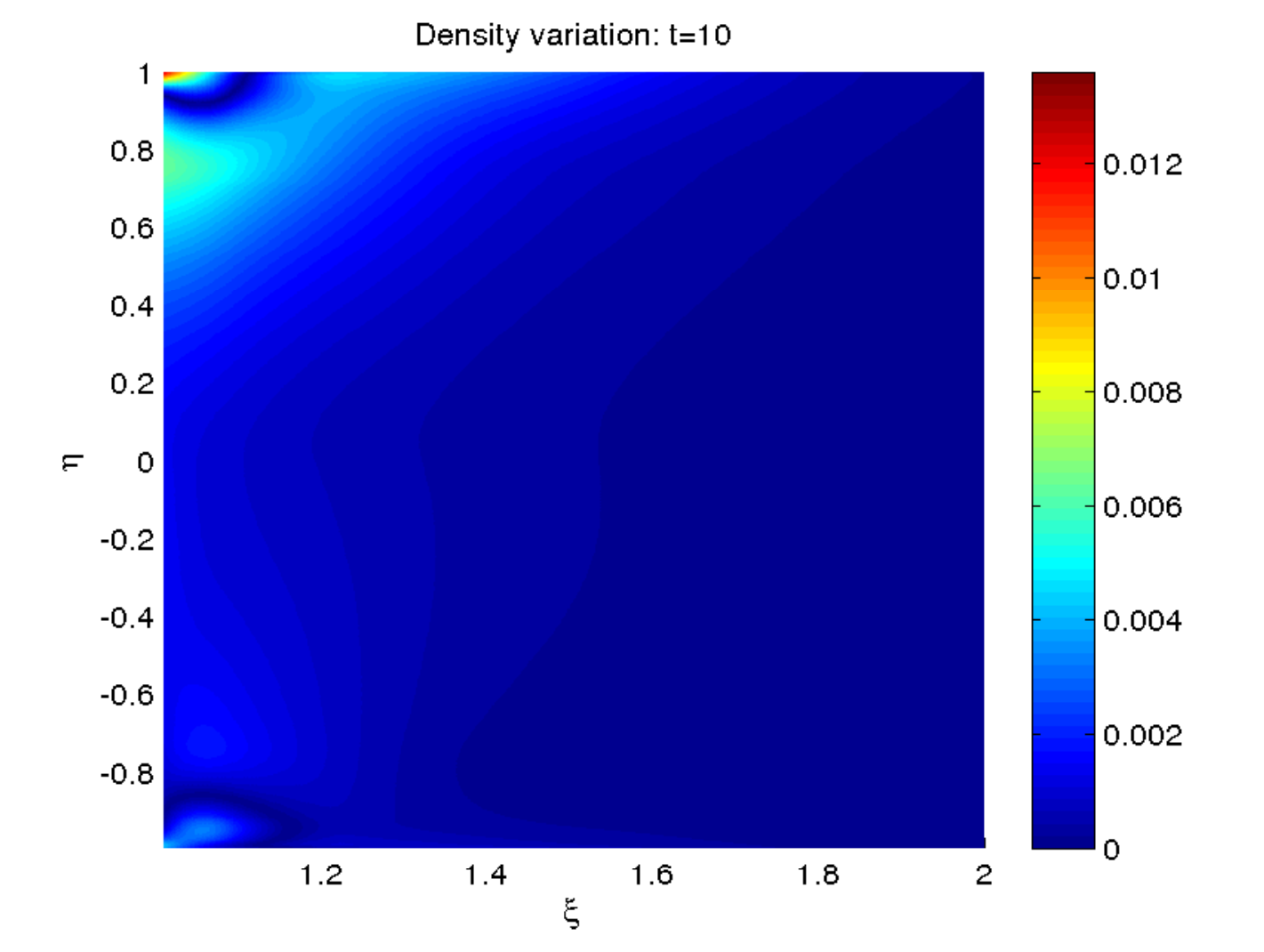}
\hspace*{1mm}\includegraphics[height=2.8cm, keepaspectratio]{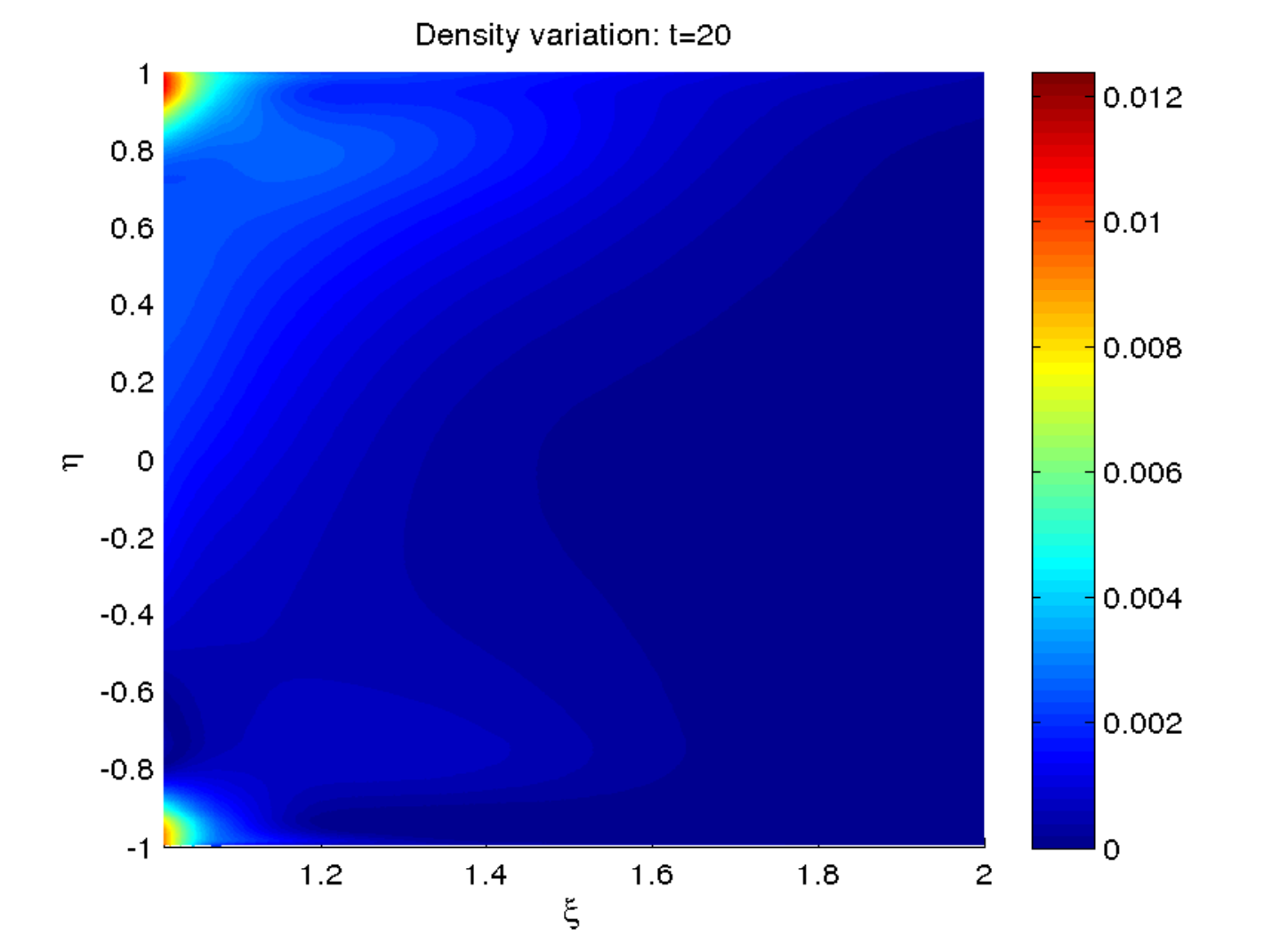}
\caption{Initial density, $\rho_0$, and Density variation $|\rho(t_f,\cdot)-\rho_0(\cdot)|$ at time $t=1$, $t=10$, $t=20$ and with $A_0=100$}
\label{field}
\end{center}
\end{figure}

One important feature of the time dependent solver is the consistency with the time independent solver (same grid, same space discretization). A first important test is then to show that without external field, the density is (almost) constant in time, as expected theoretically:
\begin{eqnarray*}
i\partial_t \psi = H_0 \psi, \, (t,x)\in \R_+\times \Omega, \qquad \psi(0,x) = \psi_0(x)
\end{eqnarray*}
with $H_0$ the field-free Hamiltonian and $\phi_0$ eigenfunction of $H_0$. The formal solution is naturally: $\psi(t,\xi,\eta)=\exp(-iH_0t)\cdot\phi_0(\xi,\eta)$ and we also have $\rho(t,\xi,\eta)=\rho_0(\xi,\eta)=\big(\sum_{i=1}^4\phi_{0,i}(\xi,\eta)^2\big)^{1/2}$, which is the initial density.  We numerically check that this consistency property is satisfied, discretizing the time derivative with a Crank-Nicolson scheme ($L^2-$norm preserving and order $2$, in time, is expected). For $8\times 8$-$(\xi,\eta)$ nodes, we first compute the $\mbox{Th}_2^{179+}$ ground state. Then using the exact same grid and spatial discretization ($N_{\xi}=N_{\eta}=8$, $N^*=182$ and B-spline order of $7$), we solve $i\partial_t \psi = H_0 \psi$. We then report in Fig. \ref{nofield}, $\big\{(\xi,\eta)\in[0,15]\times[-1,1],|\rho(t_f,\xi,\eta)-\rho_0(\xi,\eta)|\big\}$, after $10^4$ time iterations, with $\Delta t=10^{-5}$, that is $t_f=10^{-1}$. The solutions are represented on a $256\times256$ grid points. This result shows the strength of the time dependent solver with respect to its consistency with the time independent one.
\\
\\
As a preliminary example of application, we show here the interaction of a diatomic  $\mbox{H}_2^{+}$ molecule with a very short and intense external electric field polarized in the $z$-direction, for $t\geq 0$:
\begin{eqnarray*}
A_z(t)=A_0\sin^2\Big(\cfrac{\pi t}{N_0}\Big)\sin(\omega_0t)
\end{eqnarray*}
where $N_0$ is a positive integer, and $\omega_0$ the external field frequency.  We choose in atomic units, $t_f=1$, $A_0=100$, $N_0=2$ and $\omega_0=0.1$, and the numerical data are chosen as follows: $N_{\xi}=N_{\eta}=8$ and $N^*=182$, and the B-spline order is fixed at $3$. We also take $\Delta t=10^{-3}$. Note that in order to precisely describe physical phenomena up to the zitterbewegung \cite{p1}, much smaller time step is necessary ($\leq 10^{-5}$).  We report in Fig. \ref{field} the electron driven by the field, from one center to the another, at different times ($t=0,t=1,t=10,t=20$).


\section{Conclusion}\label{conc}
This paper was devoted to the derivation and analysis of a Galerkin method using atomically or kinetically B-spline basis for solving the Dirac equation. We perform spectrum, as well as time dependent evolution computations to illustrate some of the strengths of the method, such as its high order and the  consistency between the time independent and dependent solvers. It was shown that using a quite reduced number of high order B-spline basis functions, it was possible to accuratly solve the TIDE and TDDE. This is a main advantage compared to finite difference methods for instance, where a very large number of points are usually necessary for precise computations. Atomic and kinetic balance approaches were alse compared. We recalled that, in term of variational collapse, the atomic balance is more relevant than kinetic one. However, due to additional singularities, the atomic balance was shown to be harder to accurately implemented for heavy ions. A future work will be dedicated to the application of the method, to intense$\&$short laser-molecule interactions for pair production problems.

\bibliographystyle{elsarticle-num}
\bibliography{biblio}
\end{document}